\documentstyle[times,epsfig,graphics,amssymb,astrobib]{mn2e}

\newcommand{\be}{\begin{equation}}
\newcommand{\e}{\end{equation}}
\newcommand{\bear}{\begin{eqnarray}}
\newcommand{\ear}{\end{eqnarray}}

\newcommand{\de}{{\rm d}}

\title[Glimpsing through the high redshift neutral hydrogen fog]
{Glimpsing through the high redshift neutral hydrogen fog}

\author[S. Gallerani, A. Ferrara, X. Fan, T. Roy Choudhury]
{S. Gallerani$^{1}$\thanks{E-mail: galleran@sissa.it}, 
A. Ferrara$^{1}$\thanks{E-mail: ferrara@sissa.it}, 
X. Fan $^{2}$\thanks{E-mail: fan@as.arizona.edu},
T. Roy Choudhury$^{3}$\thanks{E-mail: chou@ast.cam.ac.uk} \\\\
$^1$ SISSA/International School for Advanced Studies, via Beirut
2-4, 34014 Trieste, Italy\\
$^2$ Steward Observatory, The University of Arizona, Tucson, AZ 85721, USA\\
$^3$ Institute of Astronomy, Madingley Road, Cambridge CB3 OHA, UK}

\date{\today}

\begin{document}

\maketitle

\begin{abstract}
We analyze the transmitted flux in a sample of 17 QSOs spectra at $5.74\leq z_{em}\leq 6.42$ to 
obtain tighter constraints on the volume-averaged neutral hydrogen fraction, $x_{\rm{HI}}$, at $z\approx 6$. 
We study separately the narrow transmission windows (peaks) and the wide dark portions (gaps) in the 
observed absorption spectra. By comparing the statistics of these spectral features with a semi-analytical model of the Ly$\alpha$ forest, we conclude that $x_{\rm{HI}}$ evolves smoothly from $10^{-4.4}$ at $z=5.3$ 
to $10^{-4.2}$ at $z=5.6$, with a robust upper limit $x_{\rm{HI}} < 0.36$ at $z=6.3$. The frequency and 
physical sizes of the peaks imply an origin in cosmic underdense regions and/or in HII regions around
faint quasars or galaxies. In one case (the intervening HII region of the faint quasar RD J1148+5253 
at $z=5.70$ along the LOS of SDSS J1148+5251 at $z=6.42$) the increase of the peak spectral density is explained 
by the first-ever detected transverse proximity effect in the HI Ly$\alpha$ forest; this indicates that at least some peaks result from a 
locally enhanced radiation field. We then obtain a strong lower limit on the foreground QSO lifetime of $t_Q>11$ Myr.
The observed widths of the peaks are found to be systematically larger than the simulated 
ones. Reasons for such discrepancy might reside either in the photoionization equilibrium assumption or in radiative transfer 
effects.  
\end{abstract}
\begin{figure*}
\centerline{
\psfig{figure=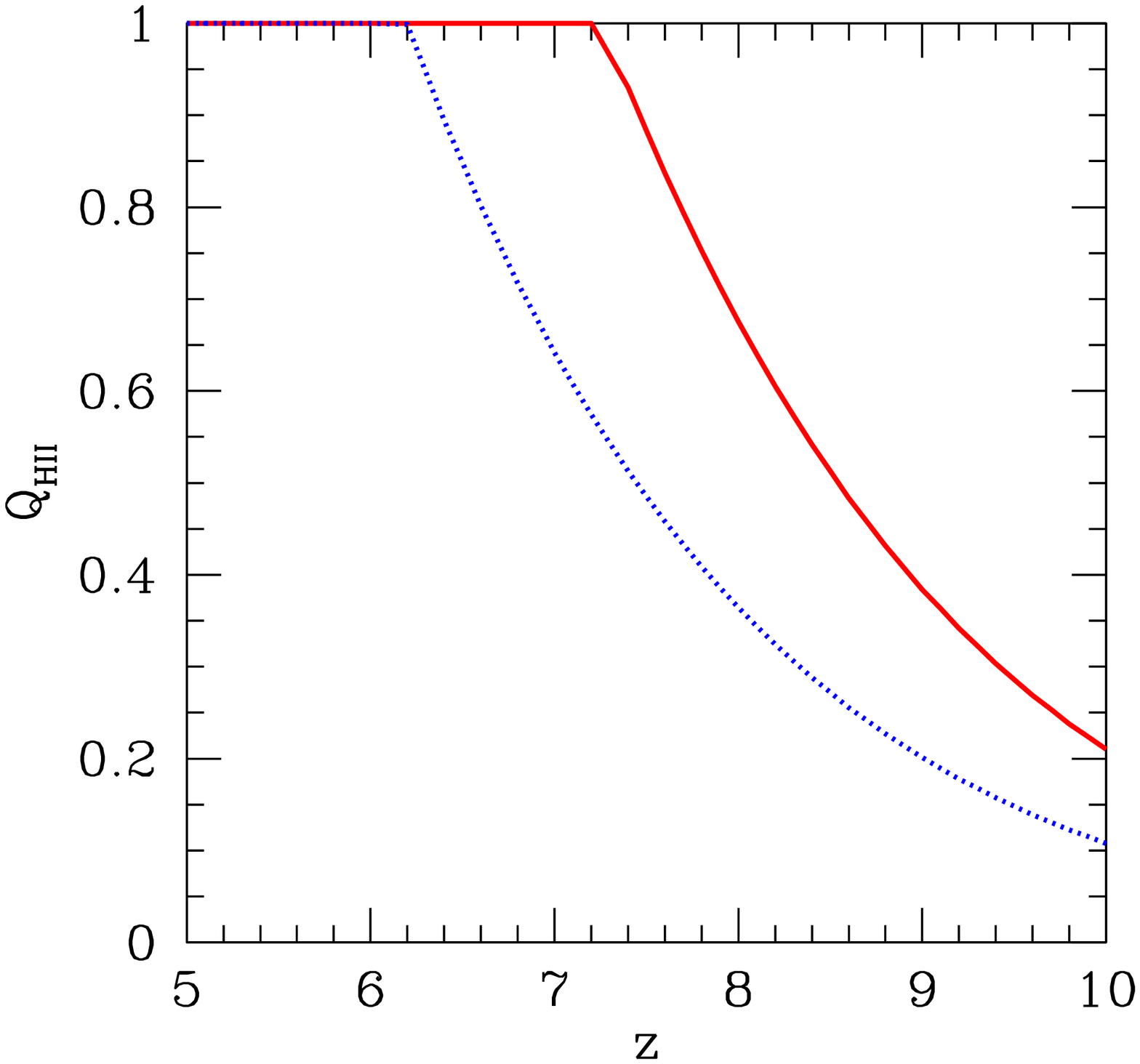,width=6.5cm,angle=0}
$\!\!\!\!\!\!\!\!\!\!\!$
\psfig{figure=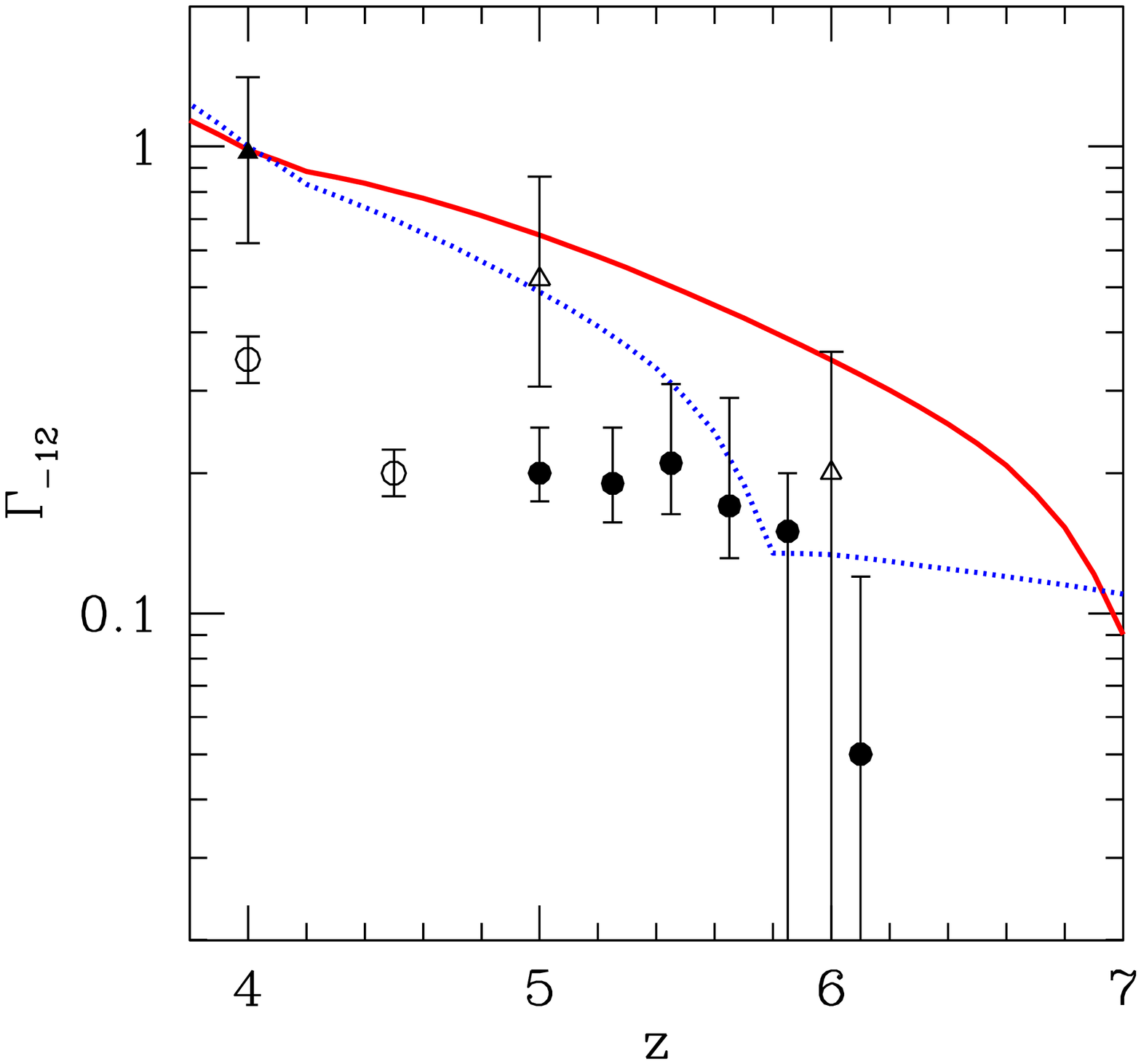,width=6.5cm,angle=0}
$\!\!\!\!\!\!\!\!\!\!\!$
\psfig{figure=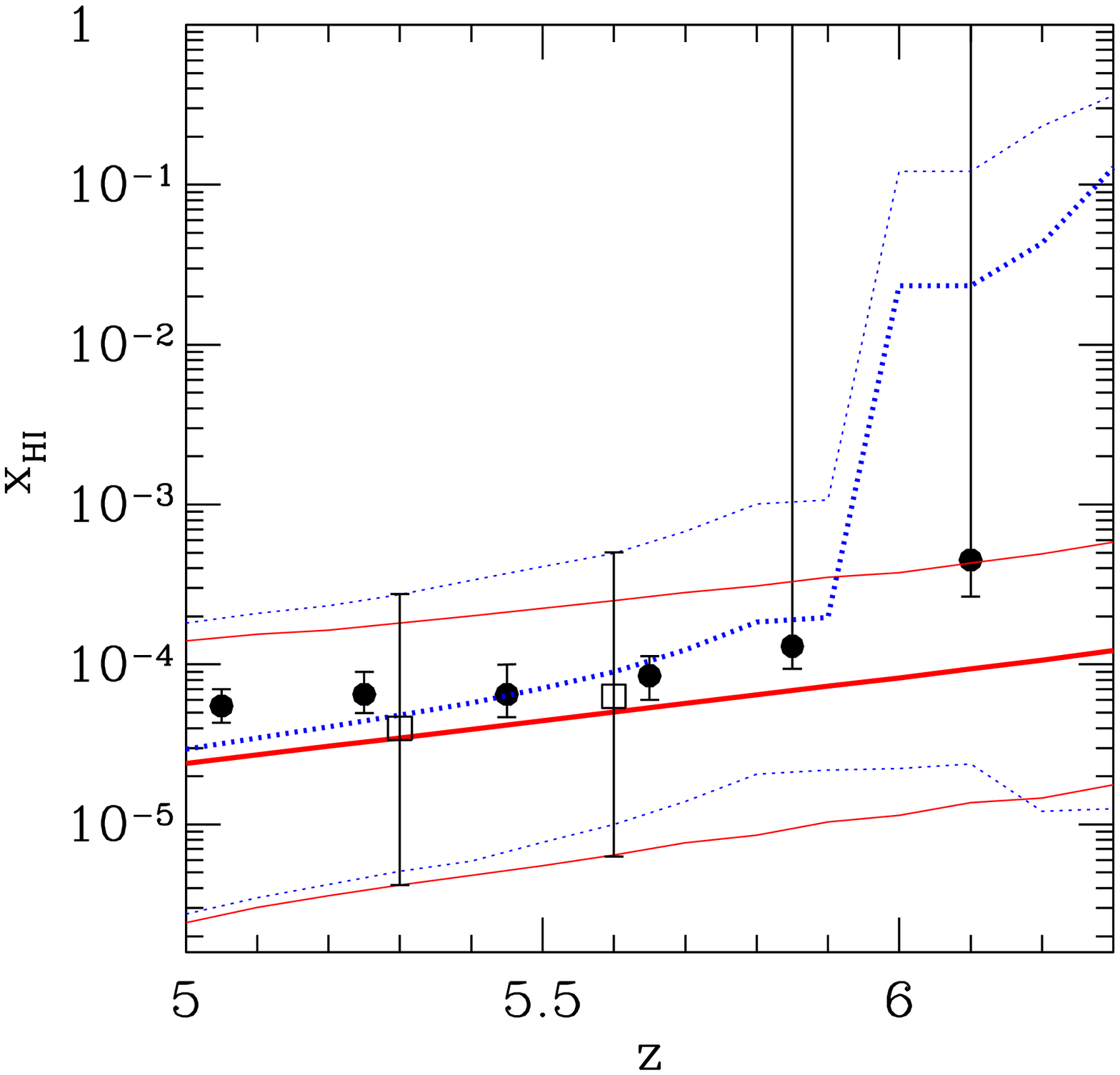,width=6.5cm,angle=0}
}
\caption{{\it Left panel}: Evolution of the volume filling 
factor of ionized regions for the early (red solid lines) and late
 (blue dotted lines) reionization models. 
{\it Middle panel}: Volume-averaged 
photoionization rate $\Gamma_{-12}=\Gamma_{\rm HI}/10^{-12}{\rm s}^{-1}$. The filled circles, empty 
circles, filled triangles and empty triangles show results obtained by F06, 
MM01, B05 and B07, respectively. {\it Right panel}: Evolution of the neutral hydrogen fraction. 
Thick lines represent average results over 100 LOS, while the thin lines 
denote the upper and lower neutral hydrogen fraction extremes in each redshift 
interval. Solid circles represent neutral hydrogen fraction estimates by F06; 
empty squares denote the results obtained in this work.}
\label{prop}
\end{figure*} 
\section{Introduction}
Although observations of cosmic 
epochs closer to the present have indisputably 
shown that the InterGalactic Medium (IGM) is in an ionized state, 
it is yet unclear when the phase transition from the neutral state to the 
ionized one started. Thus, the redshift of reionization, $z_{rei}$, is still very 
uncertain.

After the first year WMAP data a possible tension was identified between CMB 
and SDSS results. The high  electron-scattering optical depth inferred 
from the TE-EE power spectra $\tau_e\approx 0.17$ (\citeNP{ksb++03}; 
\citeNP{svp++03}) seemed 
difficult to be reconciled with the strong evolution in the Gunn-Peterson optical depth $\tau_{GP}$ 
at $z=6$ 
(\citeNP{f01}; \citeNP{f03}), accompanied by the appearance of large dark 
portions in 
QSO absorption spectra (\citeNP{bfw++01}; \citeNP{djo01}; \citeNP{f06}, hereafter F06). 
The 3-yr WMAP results have released the above tension by providing a 
smaller value for $\tau_e\approx 0.1$, which implies $z_{rei}\approx 11$ for a model with instantaneous reionization (\citeNP{page06}; \citeNP{Sp06}). However,
an increasing number of LAEs are routinely found at $z > 6$ (e.g. \citeNP{stern05}; \citeNP{iye06}; \citeNP{stark07}), possibly requiring a substantial free electron fraction, resulting in an IGM relatively transparent to Ly$\alpha$ photons.

Constraints on the IGM ionization state derived by using Ly$\alpha$ 
forest spectroscopy must take into account the extremely high sensitivity of 
$\tau_{GP}$ to tiny neutral hydrogen amounts.  Indeed, a volume averaged 
neutral hydrogen fraction as low as $x_{\rm HI}\sim 10^{-3}$ \cite{f02} is 
sufficient to completely depress the transmitted flux in QSO absorption spectra; 
thus, the detection of a Gunn-Peterson trough only translates into a lower 
limit for $x_{\rm HI}$.
For this reason, recently, many studies have tried to clarify if the SDSS data 
effectively require that the IGM was reionized as late as $z\approx 6$ 
(\citeNP{gfc06}, hereafter GCF06; \citeNP{Becker06}):
in particular, GCF06 have shown that QSO observational data currently 
available are compatible with a highly ionized Universe at that redshift.
 
Clearly the determination of the reionization epoch is strictly related to the 
measurement of the neutral hydrogen fraction at $z\approx 6$. 
To investigate this 
issue many different approaches can be used. To start with, it is worth 
mentioning that many authors have tried to constrain $x_{\rm HI}$ at high 
redshift by analyzing statistically the optical depth inside HII regions 
around high redshift QSO (\citeNP{mh04}; \citeNP{mh06}), or by measuring the 
QSO bubble sizes (\citeNP{wl04}; \citeNP{wlc05}; \citeNP{BH07}; 
\citeNP{antiesimo}, \citeNP{Lidz07}). However, sufficient ground for 
controversy remains due to intrinsic uncertainties of the various techniques.

By deriving sizes of HII bubbles surrounding observed $z=6.5$ LAEs, 
\citeNP{maro05} have provided an upper limit for the neutral hydrogen 
fraction $x_{\rm HI}\lesssim 0.2-0.5$. This result is in quite good agreement with 
the upper limit $x_{\rm HI}\lesssim 0.45$ found by \citeNP{kashi06}, 
by interpreting the deficit measured at the bright end of the LAE Luminosity Function at
$z>6$ as a sudden change in the intergalactic neutral hydrogen content. 
Nevertheless, the increasing attenuation with redshift of the Ly$\alpha$ line 
transmission could be partially explained as a consequence of the 
evolution in the mass function of dark matter halos, thus implying a much 
lower upper limit,  $x_{\rm HI}<0.05-0.2$ \cite{dij07}.
 
GRB spectroscopy has also tentatively used to constrain $x_{\rm HI}$; 
\citeNP{totani06} 
have observed a damping wing at wavelengths larger than the
Ly$\alpha$ emission line, finding that this feature can be explained at best 
by assuming an intervening damped Ly$\alpha$ system immersed in a fully 
ionized IGM, and quoting an upper limit of $x_{\rm HI}<0.17$ and $0.60$ ($68\%$ and $95\%$ confidence levels, respectively).

Finally, the width distribution of dark portions (gaps) seen in QSO absorption spectra 
has been recently introduced in order to constrain the IGM ionization state (\citeNP{pn05}; F06; GCF06). 
F06 has used the dark gap distribution, as observed in 19 high-$z$ QSO spectra, to put a preliminary  upper limit on 
the IGM neutral fraction $x_{\rm HI}<0.1-0.5$.  GCF06, by analyzing the statistically properties of the transmitted flux 
in simulated absorption spectra, have shown that the gap and peak (i.e. transmission windows) 
width statistics are very promising tools for discriminating between an 
early ($z_{rei} > 6$) and a late ($z_{rei}\approx 6$) reionization scenario. 
Here we combine the previous two results: by comparing the observed 
transmitted flux in high-$z$ QSO spectra with theoretical predictions we 
obtain tighter constraints on the neutral hydrogen fraction around $z=6$, a 
crucial epoch in the reionization history. 

The plan of the paper is the following: in Section 2 we describe the 
semi-analytical modeling adopted; in Section 3 we compare observational data 
with simulations. The implications of this comparison are given in Section 
4; in Section 5 we evaluate the robustness of our method against the 
specific line of sight to the highest redshift QSO.  The conclusions are summarized
in Section 6. 

\section{Simulations}
The radiation emitted by QSOs could be absorbed through Ly$\alpha$ transition 
by the neutral hydrogen intersecting the line of sight, the so-called 
Gunn-Peterson (GP) effect. 
The Ly$\alpha$ forest arises from absorption by low amplitude-fluctuations in 
the underlying 
baryonic density field \cite{bbc92}, and is a natural consequence of the 
hierarchical 
structure formation expected in the context of CDM cosmologies\footnote
{Throughout this 
paper we will assume a flat universe with total matter, vacuum, and baryonic 
densities in units of 
the critical density of $\Omega_m=0.24$, $\Omega_{\Lambda}=0.76$, and 
$\Omega_bh^2=0.022$, respectively, and a Hubble constant of 
$H_0=100 h$~km~s$^{-1}$Mpc$^{-1}$, with $h=0.73$. The parameters defining the 
linear dark matter power spectrum are $n=0.95$, 
$\de n/\de \ln k=0$, $\sigma_{8}=0.82$. Note that we have chosen 
a $\sigma_8$ value higher than the WMAP3 one ($0.74$). 
Indeed \citeNP{viel06}, by combining WMAP3 data with SDSS ones, 
found $\sigma_8\approx 0.78$ ($0.86$) analyzing high (low) 
resolution Ly$\alpha$ forest data.
Mpc are physical unless differently stated.}. 

To simulate the GP optical depth ($\tau_{GP}$) distribution we use the method 
described by GCF06, whose
main features are recalled in the following. The spatial distribution of the 
baryonic density field and its correlation with the peculiar velocity field are taken into account adopting the formalism introduced by \citeNP{bd97}. To enter the mildly non-linear regime 
which characterizes the Ly$\alpha$ forest  absorbers we use a Log-Normal 
model introduced by \citeNP{cj91}, widely adopted later 
on (\citeNP{bi93}; \citeNP{bd97}; \citeNP{cps01}; \citeNP{csp01}; 
\citeNP{vmht02}; GCF06). In particular, GCF06 have compared various Ly$\alpha$ statistics, namely the Probability Distribution Function (PDF) and the Gap Width distribution, computed using the Log-Normal distribution with those obtained from \rm{HYDROPM} simulations, finding a good agreement between the results.
For a given IGM temperature, the HI fraction, $x_{\rm HI}$, can be computed from the
photoionization equilibrium as a function of the baryonic density field and photoionization rate 
due to the ultraviolet background radiation field. For all these quantities we 
follow the approach of \citeNP{cf06}, hereafter CF06.  By assuming as ionizing sources QSOs, PopII and PopIII stars (the latter neglected here,
see below), 
their model provides excellent fits to a large number of observational data, namely the redshift evolution of 
Lyman-limit systems, Ly$\alpha$ and Ly$\beta$ optical depths, electron scattering optical depth, cosmic star formation history, and the number counts of high redshift sources.
\begin{figure*}
\centerline{
\psfig{figure=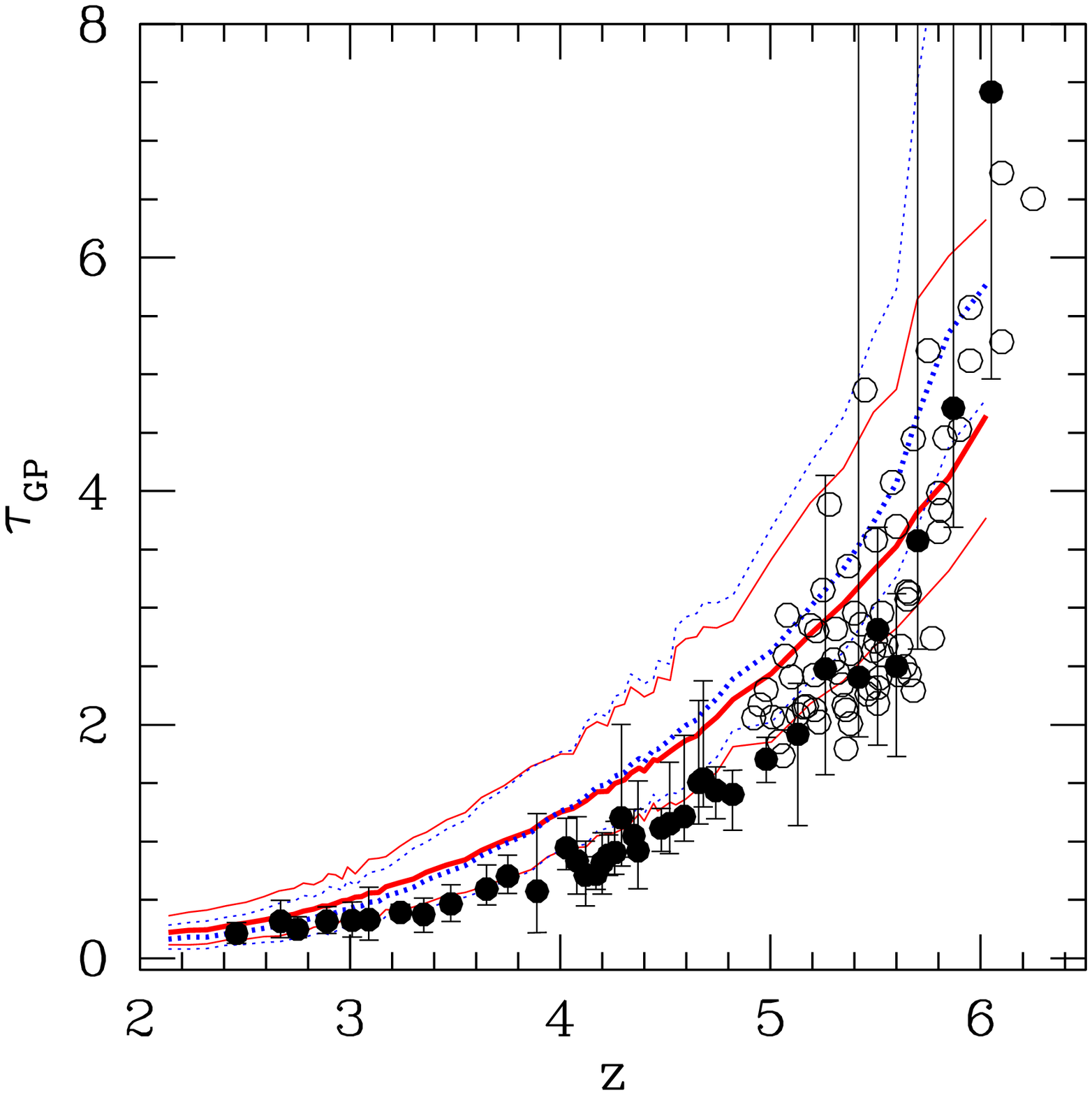,width=6.5cm,angle=0}
$\!\!\!\!\!\!\!\!\!\!\!$
\psfig{figure=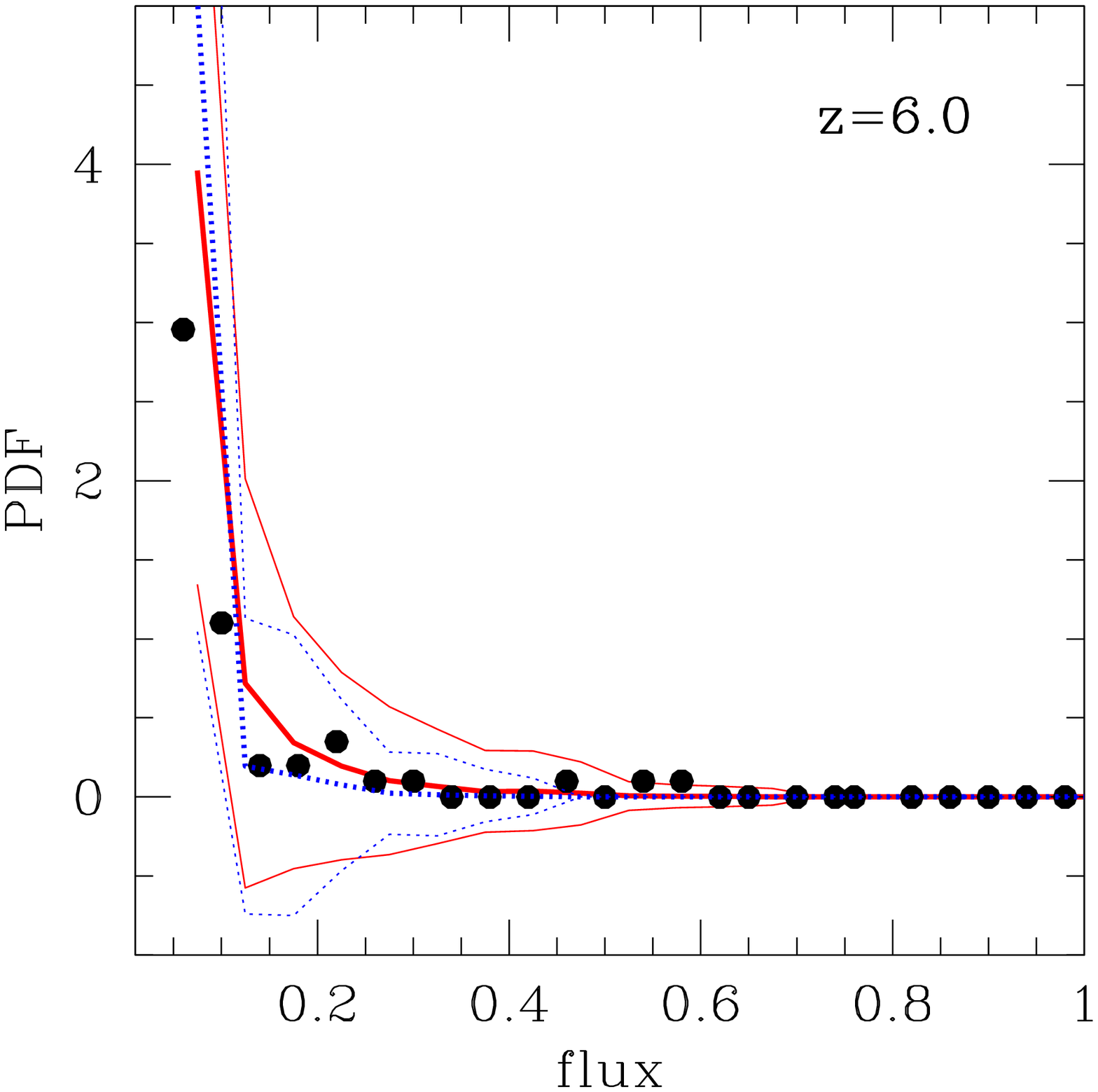,width=6.5cm,angle=0}
$\!\!\!\!\!\!\!\!\!\!\!$
\psfig{figure=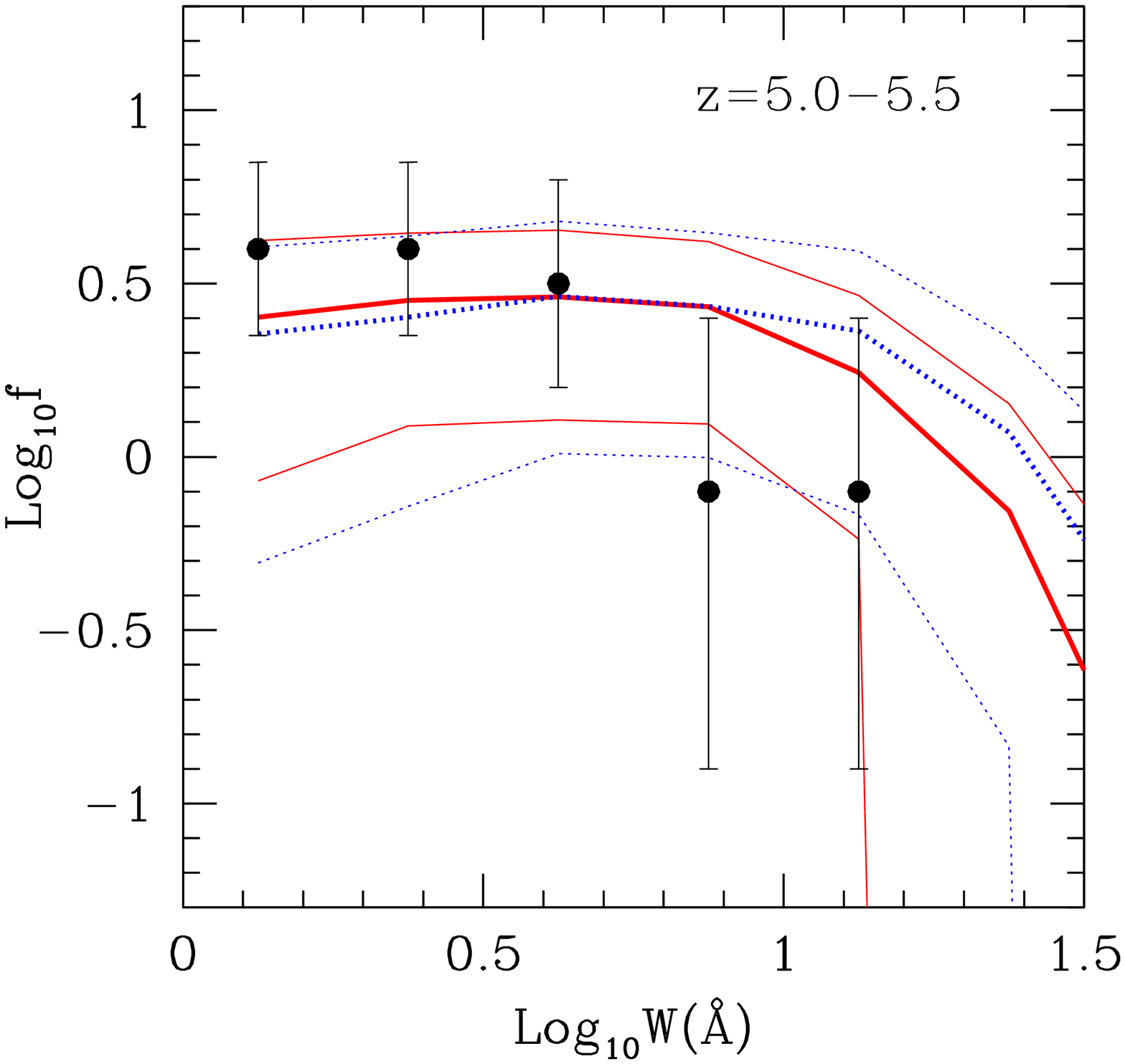,width=6.5cm,angle=0}
}
\caption{{\it Left panel}: Evolution of the Gunn-Peterson 
optical depth for early (ERM, solid red line)  and late (LRM, blue dotted). 
Thick lines represent average results on 100 LOS for each emission redshift, 
while the thin lines denote the upper and lower transmission extremes in 
each redshift bin, weighted on 100 LOS. Filled and empty circles are 
observational 
data from Songaila 2004 and F06, respectively. {\it Middle panel}: Probability 
Distribution Function (PDF) of the flux at z=6.0. Filled circles are obtained 
by Fan et al. 2002. Thick lines represent simulated results averaged over 500 
LOS, while the thin lines denote cosmic variance. {\it Right panel}: Gap Width 
distribution in the redshift range 5.0-5.5. Simulated results are compared 
with observations by Songaila \& Cowie 2002 (filled circles). 
The errors associated to both simulated and observed results denote cosmic 
variance.}
\label{controlfig}
\end{figure*} 
In the CF06 model, a reionization scenario is defined by the product of two free 
parameters: (i) the star-formation efficiency $f_*$, and (ii) the escape fraction $f_{esc}$ 
of ionizing photons of PopII and PopIII stars; it is worth noting that these parameters are degenerate, 
since different parameter values could provide equally good fits to observations.
In this work, by fitting all the above observational constraints, we select two sets of free parameters values yielding two different 
reionization histories: (i) an Early Reionization Model (ERM) for 
($f_{*,PopII}=0.1; f_{esc,PopII}=0.07$), and (ii) a Late Reionization Model (LRM) for ($f_{*,PopII}=0.08; f_{esc,PopII}=0.04$). 
We do not consider contributions from PopIII stars, as PopII stars alone yield $\tau_e=0.07$ ($0.06$) for ERM (LRM), marginally consistent with 
WMAP3 results\footnote{Small contributions from PopIII stars, i.e. 
$f_{*,PopIII}=0.013$ ($f_{*,PopIII}=0.08$), in the ERM (LRM), would yield 
$\tau_e=0.09$ ($\tau_e=0.08$), without affecting sensitively the results below.}.\\
Fig. \ref{prop} shows the global properties of the two reionization models 
considered.  In the ERM the volume filling factor of ionized regions, 
$Q_{HII}=V_{HII}/V_{tot}=1$ at $z\leq 7$; in the LRM it evolves from 0.65 to 
unity in the redshift range 7.0-6.0, implying that the Universe is still in the 
pre-overlap stage at $z\geq 6$, i.e. the reionization process is not completed up to this epoch. In the middle panel of the same Figure we compare the volume-averaged 
photoionization rate $\Gamma_{\rm HI}$ for the two models with 
the recent estimate by F06, and the ones by \citeNP{mm01}, 
\citeNP{BH05}, and \citeNP{BH07new}, hereafter MM01, B05 and B07, respectively. Finally, the evolution of the 
volume-averaged neutral hydrogen fraction for the ERM and LRM is presented in the rightmost panel. 

The photoionization rate predicted by both models is in agreement with the
results by B05 and B07 at in the range $z=4.0 < z < 6$, whereas at $z=5.5$ ($6$) 
the ERM is characterized by a photoionization rate which is $\approx$ 2 (6) times larger than the 
estimates by F06. In spite of these differences, our predictions for $x_{\rm HI}$ are consistent 
with F06 measurements. This apparent contradiction does not come as a 
surprise. In fact, the derivation of $\Gamma_{\rm HI}$ requires an assumption 
concerning the IGM density distribution. When measuring $\Gamma_{\rm HI}$ at 
$5<z<6$, F06 assume the density Probability Distribution Function given by 
\citeNP{mhr00}, hereafter MHR\footnote{F06 require $\Gamma_{\rm HI}$ to match 
the MM01 measurement at $z=4.5$. 
This estimate is based on a mean transmitted flux ($\bar{F}=0.25$) which is 
lower than the more recent measurements $\bar{F}\approx 0.32$ by 
\citeNP{songaila04}, which implies $\Gamma_{\rm HI}\approx 0.3$.}.
We instead adopt a Log-Normal 
(LN) model which predicts a higher probability to find overdensities 
$\Delta=\rho/\bar{\rho}\gtrsim 1$ than MHR. For example, at $z=6$ and for $\Delta\approx 1.5$, 
$P_{LN}(\Delta)\approx 2\times P_{MHR}(\Delta))$. For this reason, 
once $\tau_{GP}$ is fixed to the observed value, the LN model requires a 
higher $\Gamma_{\rm HI}$. As $x_{\rm HI}\propto \Delta$, these two effects combine
to give a values of $x_{\rm HI}$ consistent with the data. 
\begin{figure*}
\psfig{figure=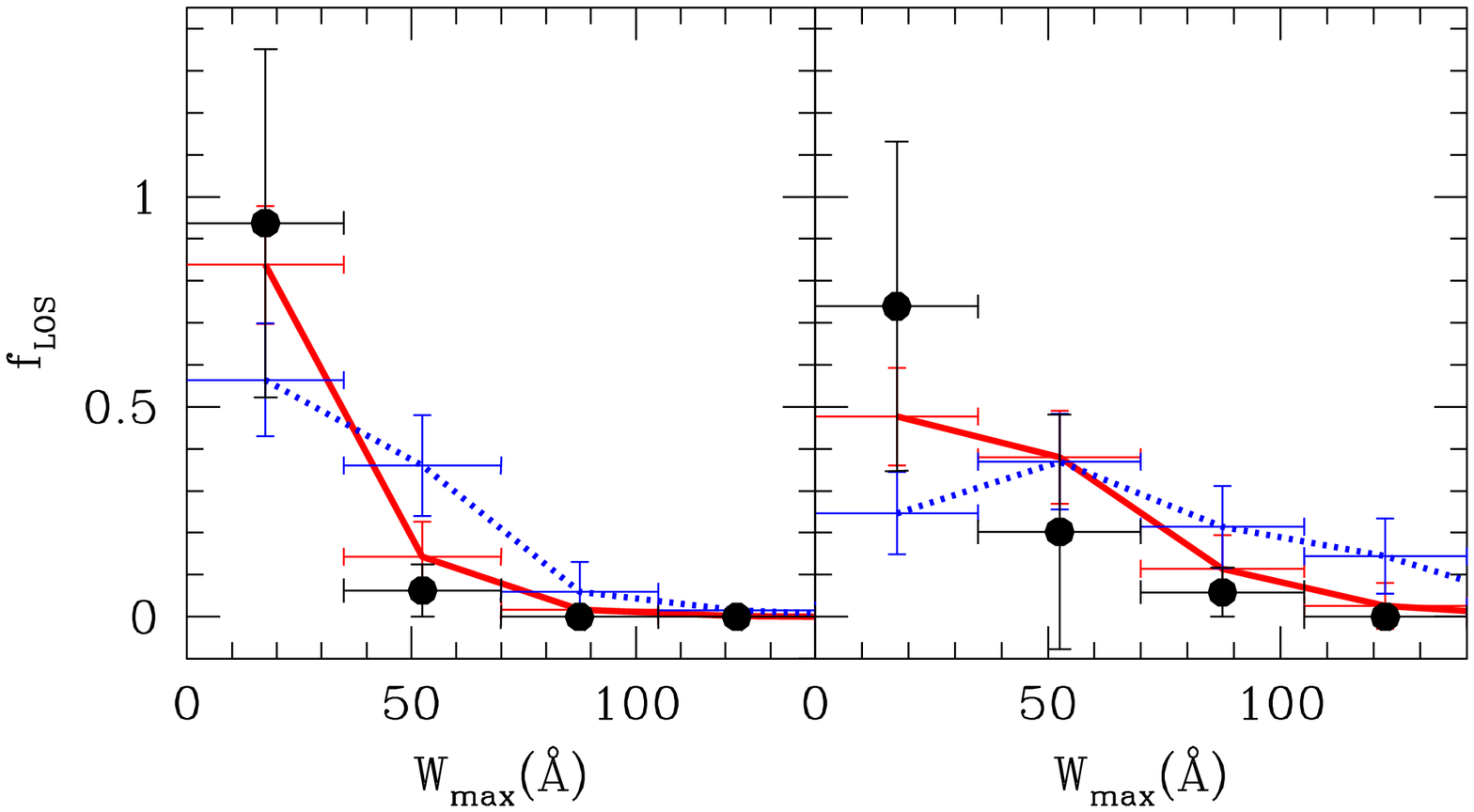,height=8.cm}
\caption{Largest Gap Width distribution for the LR and the 
HR cases (left and right, respectively). 
Filled circles represent the result of the analysis of the 17 QSOs observed 
spectra.
Solid red (dotted blue) lines show the 
results obtained by the semi-analytical modeling implemented for the ERM 
(LRM). Vertical error bars measure poissonian noise, horizontal errors define 
the bin for the gap widths.}
\label{larg}
\end{figure*}
\section{Comparison with observations}
\subsection{Control statistics}
We first test the predictions of our model by applying various
statistical analysis to the simulated spectra and comparing our results with 
observations. Specifically, we use the following control statistics:
(i) Mean Transmitted Flux evolution in the redshift range $2-6$; 
(ii) Probability Distribution Function (PDF) of the transmitted flux 
at the mean redshifts $z=5.5, 5.7, 6.0$; (iii) Gap Width (GW) distribution in 
$3.5\leq z\leq 5.5$. For what concerns the GW statistics we define gaps as 
contiguous regions of the spectrum having a $\tau_{GP}> 2.5$ 
over rest-frame wavelength ($\lambda_{RF}$) intervals $> 1$~\AA.  
This method was first suggested by Croft (1998) and then adopted by various 
authors (\citeNP{sc02}; \citeNP{pn05}; F06; GCF06). The comparison
of model and observational results in terms of the above three statistics 
is plotted in Fig. \ref{controlfig}. By checking our models we 
follow the same approach of GCF06, to which we refer for a complete 
description of the technical details. 

The outcome of the test is encouraging, as both ERM and LRM successfully match 
the observational data at $z\leq 6$ for the control statistics considered. 
This allows us to confidently proceed the comparison with more advanced 
statistical tools. 

\subsection{Advanced statistics}
Since at $z\approx 6$ regions with high transmission in the Ly$\alpha$ forest 
become rare, an appropriate method to analyze the statistical properties 
of the transmitted flux is the distribution of gaps. In particular 
GCF06 suggested that the Largest Gap Width (LGW) and the Largest Peak Width 
(LPW) 
statistics are suitable tools to study the ionization state 
of the IGM at high redshift\footnote{The definition of ``peak'' in the 
transmitted flux is similar to the ``gap'' one. A peak is a contiguous 
region of the spectrum over $\lambda_{RF}$ intervals greater than the observed pixel size ($\approx 0.5$ \AA) characterized by a transmission above a given flux threshold 
($F_{th}=0.08$ in this work).}.
The LGW (LPW) distribution quantifies the fraction of LOS 
which are characterized by the largest gap (peak) of a given width. 
As far as this work is concerned, we apply the LGW and the LPW statistics both 
to simulated and observed spectra with the aim of measuring  the evolution of 
$x_{\rm HI}$ with redshift. \\
We use observational data including 17 QSOs obtained by F06.
We divide the observed spectra into two redshift-selected sub-samples: 
the ``Low-Redshift'' (LR) sample (8 emission redshifts $5.7 < z_{em} < 6$), and
the ``High-Redshift'' (HR) one (9 emission redshifts $ 6 < z_{em} < 6.4$). 
Simulated spectra  have the same $z_{em}$ distribution of the observed samples.
For most QSOs we consider the ($\lambda_{RF}$) interval 1026-1200 \AA\ and we normalize each 
width to the corresponding redshift path. Note that the LOS do not extend up to $z_{em}$; 
the upper (lower) limit of the interval chosen ensures that we exclude from the analysis 
the portions of the spectra penetrating inside the QSO HII (Ly$\beta$) region. 
For the QSOs SDSS J1044-0125 and SDSS J1048+4637 we 
choose different intervals, namely 1050-1183 and 1050-1140, respectively. 
These two objects have been classified as 
BAL QSO (\citeNP{goodrich}; \citeNP{f03}; 
\citeNP{maio04}), since their spectra present Broad Absorption 
Lines associated with highly ionized atomic species (e.g. SiIV, CIV).
By selecting the above intervals we exclude those portions of the spectra 
characterized by CIV absorption features which extend to 
$z\approx 5.56$ ($z\approx 5.75$) in SDSS J1044-0125 (SDSS J1048+4637).\\
Observed data were taken with a spectral resolution $R\approx 3000-6000$; 
simulated spectra have been convolved with a gaussian of 
$FWHM=67~{\rm km/sec}$, providing $R\sim 4500$. Moreover each 
observed/simulated spectrum has been rebinned to a resolution of $R=2600$. 
Finally, we add noise to the simulated data such that the flux $F$ in each pixel is replaced by 
$F+G(1)\times\sigma_n$, where $G(1)$ is a Gaussian random deviate with zero 
mean and unit variance, and $\sigma_n$ is the observed noise r.m.s deviation 
of the corresponding pixel. 

The results provided by the statistics adopted in 
this study are sensitive to the S/N ratio, since spurious peaks 
could arise in spectral regions with noise higher than the $F_{th}$ adopted. 
Indeed, the shape of the LGW/LPW distributions depends on the $F_{th}$ 
chosen. Thus, we consider two 
different values for $F_{th}$, namely $0.03$ and $0.08$, respectively, and, 
for both of them, compute preliminary LGW/LPW distributions. 
Finally, the  LGW/LPW distributions presented are obtained as the mean of the 
preliminary ones, weighted on the corresponding errors 
(See Appendix A for a detailed discussion).
In our analysis we do not consider 2 QSOs presented by F06, 
namely SDSS J1436+5007 and SDSS J1630+4012, since these spectra have 
significantly lower S/N to apply LGW/LPW tests (continuum S/N $\lesssim 7$).
\begin{figure*}
\psfig{figure=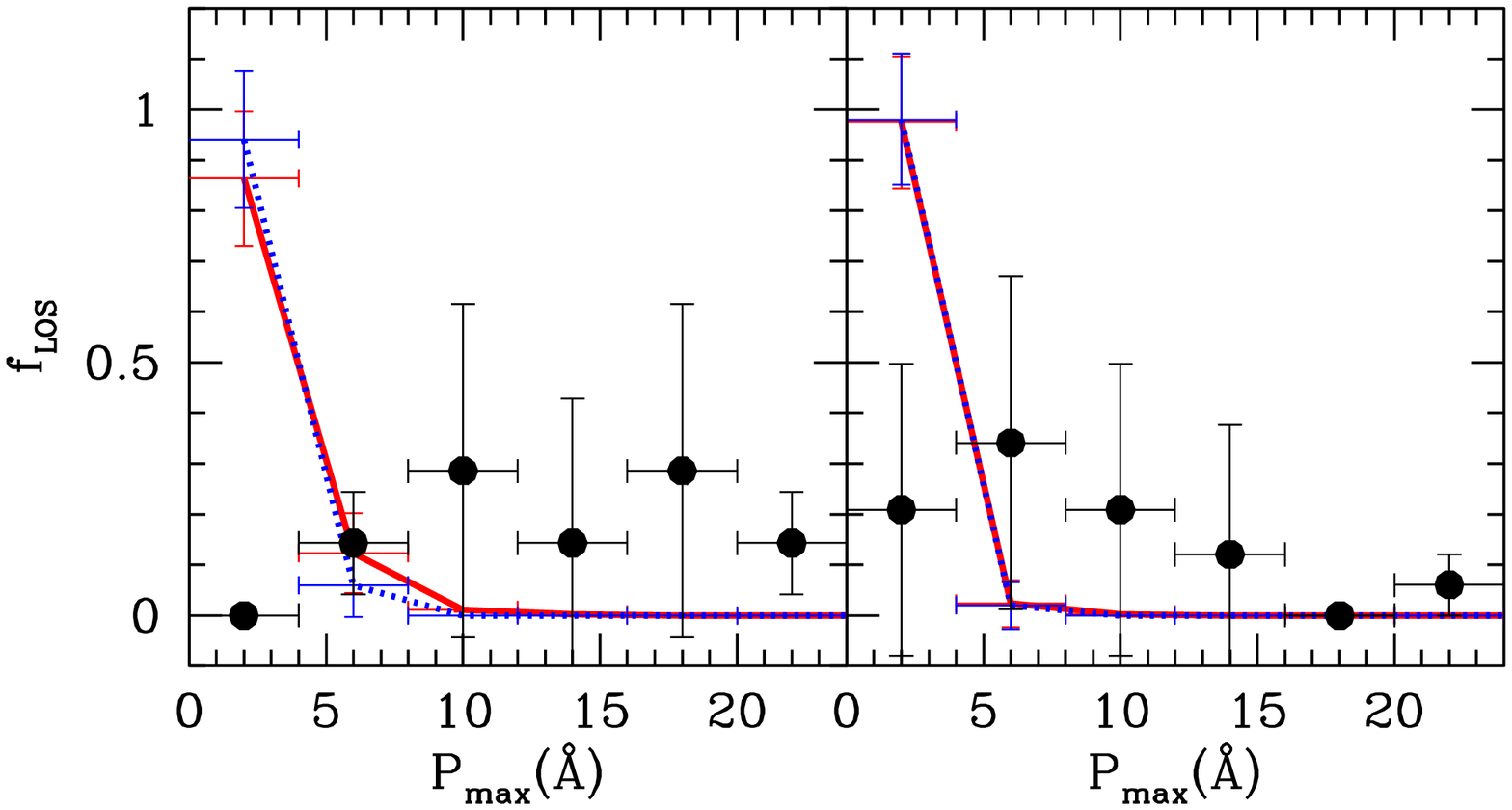,height=8.cm}
\caption{Largest Peak Width distribution for the LR and the 
HR cases (left and right, respectively). 
Filled circles represent observational data obtained by analyzing the observed 
spectra of the 17 QSOs considered. 
Solid red (dotted blue) lines show the 
results obtained by the semi-analytical modeling implemented for the ERM 
(LRM). Vertical error bars measure poissonian noise, horizontal errors define 
the bin for the peak widths.}
\label{peakfig}
\end{figure*}
\subsubsection{\it LGW distribution}
We now discuss the LGW distribution for observed/simulated spectra; 
the results are shown in Fig. \ref{larg}.
The QSOs emission redshifts used and the $\lambda_{RF}$ 
interval chosen for the LR sample are such that the mean redshift of the 
absorbers is $\langle z\rangle=5.26$, with a minimum (maximum) redshift $z_{min}=4.69$ 
($z_{max}=5.86$), and a r.m.s. deviation $\sigma=0.06$. For the HR sample it is $\langle z\rangle=5.55$, $z_{min}=4.90$, $z_{max}=6.32$, 
$\sigma=0.14$. 
The observed LGW distribution evolves rapidly with redshift: in the LR sample most of the LOS are 
characterized by a largest gap $<40$~\AA, whereas gaps as 
large as 100 \AA \ appear in the HR sample. This means that LOS to QSOs 
emitting at $z_{em}\lesssim 6$ encounter ``optically thick'' regions whose size is 
$\leq$ 20 Mpc, 
while for $z_{em}\gtrsim 6$ blank regions of size up to 46 Mpc are present.\\
Superposed to the data in Fig. \ref{larg} are the predicted LGW distributions 
corresponding to ERM and LRM, obtained by simulating 800/900 LOS in the LR/HR case, corresponding to 100 LOS for each emission redshift in each sample. 
In our ERM simulated spectra, at $z\approx 6$ gaps are produced by 
regions characterized by a mean overdensity $\bar{\Delta}\approx 1$ ($\Delta_{min}=0.05$, $\Delta_{max}=18$) with a 
$x_{\rm HI}\approx 10^{-4}$, averaging on 100 LOS ($x_{\rm HI,min}=1.1\times 10^{-5}$, 
$x_{\rm HI,max}=3.6\times 10^{-4}$). \\
It results that both the predicted LGW distributions provide a good fit to 
observational data. We exploit the agreement between the simulated and 
observed LGW distributions to derive an estimate of $x_{\rm HI}$, shown in 
Fig. \ref{prop}. We find $\log_{10}x_{\rm HI}=-4.4^{+0.84}_{-0.90}$ at $z\approx 5.3$\footnote{The $x_{\rm HI}$ value 
quoted is the mean between the estimates predicted by the ERM and the LRM. 
Moreover, we consider the most conservative case in which the errors for the 
measurement of the neutral hydrogen fraction are provided by the minimum 
$x_{\rm HI}$ value found in the ERM and the maximum one in the LRM.}.
By applying the same method to the HR sample we constrain the 
neutral hydrogen fraction 
to be within $\log_{10}x_{\rm HI}=-4.2^{+0.84}_{-1.0}$ at $z\approx 5.6$.\\ 
Although the predicted LGW distributions are quite similar for the two models 
considered, yet some differences can be pointed out.
Both for the LR and HR cases the early reionization LGW 
distribution provides a very good match to the observed points, thus suggesting $z_{rei}\gtrsim 7$. 
The agreement is satisfactory also for the LRM, but it is important 
to notice that late reionization models predict too 
many largest gaps $\approx$ 60 \AA \ in the LR case and too few gaps $\approx$ 20 
\AA \ in the HR one. Given the 
limited quasar sample available, the statistical relevance of the LRM 
discrepancies is not sufficient to firmly rule out this scenario. However, 
 since in the HR case $40\%$ of the lines of sight extend at $z\gtrsim 6$, we can use the LRM results to put an upper 
limit on $x_{\rm HI}$ at this epoch.
Indeed in the HR case we find that a 
neutral hydrogen fraction at $z\approx 6$ 
higher than that one predicted by the LRM would imply an even worst 
agreement with observations, since a more abundant HI would produce a lower 
(higher) fraction of LOS characterized by the largest gap smaller (higher) 
than $40$ \AA\ with respect to observations. 
Thus, this study suggests $x_{\rm HI}<0.36$ at $z=6.32$ (obtained from the 
maximum value for $x_{\rm HI}$ found in the LRM at this epoch).
\subsubsection{\it LPW distribution}
Next, we apply the Largest Peak Width (LPW) statistics (Fig \ref{peakfig}) to 
both observed and simulated spectra. 
From the observed LPW distribution we find that, in the LR (HR) sample, 
about $50\%$ of the lines of sight exhibit peaks of width $< 12 (8)$ \AA.
In more details, the size $P_{max}$ of the largest transmission 
regions in the observed sample are $3\lesssim P_{max}\lesssim 10$ 
($1\lesssim P_{max}\lesssim 6$) Mpc at $\langle z\rangle=5.3$ ($5.6$).
The frequency and the 
amplitude of the transmission regions rapidly decrease toward high redshift. 
This could be due both 
to the enhancement of the neutral hydrogen abundance at epochs approaching 
reionization or 
to evolutionary effects of the density field \cite{songaila04}. In fact the growth factor $D_+$ 
of density fluctuations decreases with redshift ($D_+(z=6)=3/5\times D_+(z=3)$ 
for $\Lambda CDM$), thus implying a low density contrast at $z=6$ with respect 
to later 
epochs. Stated differently, underdense regions that are transparent 
at $z=3$, were less underdense at $z=6$, thus blocking transmission. 
As a consequence of the density field evolution toward higher $z$, only few/small 
peaks survive and wide GP troughs appear. 

Superposed to the data in Fig. \ref{peakfig} are the predicted LPW distributions 
corresponding to ERM and LRM, obtained by simulating 800/900 LOS in the LR/HR case. 
In our ERM simulated spectra, at $z\approx 6$, gaps are interrupted by 
narrow transparent windows (i.e. peaks) originating from underdense regions 
with $\bar{\Delta}\approx 0.1$, averaging on 100 LOS ($\Delta_{min}=0.03$, $\Delta_{max}=0.26$) and 
$x_{\rm HI}\approx 2\times 10^{-5}$, ($x_{\rm HI,min}=7.8\times 10^{-6}$, 
$x_{\rm HI,max}=3.6\times 10^{-5}$). Regions characterized by 
$\Delta\in [0.05;0.26]$ and 
$x_{\rm HI}\in[1.1\times 10^{-5};3.6\times 10^{-5}]$ 
could correspond to both gaps or peaks depending on redshift and  
peculiar motions of the absorbers producing them.

By comparing the simulated LPW distributions with the observed one, 
it is evident that simulations predict peak widths that are much smaller than the observed ones
both for LR and HR cases. In particular, in no LOS of our simulated samples 
we find peaks larger than 8 \AA.  The disagreement between the 
observed and simulated LPW distributions does not affect the estimate of 
$x_{\rm HI}$ through the LGW distributions, since 
at high redshift the peaks are narrow  ($\lesssim 10$~\AA).
We discuss the possible reasons for this discrepancy in the final Section. 

\section{PHYSICAL INTERPRETATION OF THE PEAKS}
The most natural interpretation for the peaks is that they correspond to underdense regions, where
the low HI density of the gas allows a high transmissivity. However, in principle they could also arise
if individual ionized bubbles produced by QSOs and/or galaxies are crossed by the LOS. In the
latter case the typical physical size and frequency of such semi-transparent regions must be
related to the emission properties and masses of such objects. Stated differently, 
the fraction of LOS, $f_{LOS}$, having the largest peak width equal to $P_{max}$ can be 
interpreted as the probability $\wp$ to intersect an HII region of radius $R_{HII}$ around a
dark matter halo hosting either a QSO or a galaxy along the redshift path ($z_i-z_f$) 
spanned by the LOS. 
The comoving number density $n_h$ of dark matter halos of mass $M_h$ is related to $\wp$ 
through the following equation:
\begin{equation}
n_h(M_h)=\frac{3}{2} \frac{H_0 \Omega_m^{1/2}}{c} (\pi R_{HII}^2)^{-1}\left[\left(1+z\right)^{3/2}|^{z_f}_{z_i}\right]^{-1}\wp,
\label{probhalo}
\end{equation} 
We take $R_{HII}=1 (10)$ Mpc, consistent with the smaller (larger) size 
$P_{max}$ of the observed 
largest peaks in the HR (LR) sample. As it is likely that statistically the 
LOS crosses the 
bubble with non-zero impact parameter, adopting $R_{HII}=P_{max}$ seems a 
reasonable assumption.
By further imposing $\wp=f_{LOS}$ we find that $n_h=3.7\times 10^{-6}$ 
($2.2\times 10^{-8}$) Mpc$^{-3}$
for $P_{max}=1$ ($10$) Mpc in the redshift range $z_i=5$ to $z_f=6$. Given our 
cosmology, such 
halo number density can be transformed at $z=5.5$ into a typical halo mass of 
$M_h\gtrsim 10^{12}$ 
($10^{13}$) ${\rm M}_{\odot}$ \cite{mowhite}. Thus, the halos hosting the 
putative luminous sources
producing the peaks must be massive. Note that this result holds even if 
the QSO is shining only for a fraction of the Hubble time $t_{Q}/t_H \approx 10^{-2}$ 
at $z=5.5$. \\
In addition to the peak frequency, additional constraints on the properties of 
the ionizing sources come from bubble physical sizes.  
\subsection{QSO HII regions}
First, we consider the case in which the largest peaks are produced by HII regions around QSOs.
The bubble size $R_{HII}$ is related to the ionizing photons emission rate $\dot{N}_{\gamma}$ 
and QSO lifetime $t_Q$ as
\begin{equation}\label{rh2}
R_{HII}=\left( \frac{3\dot{N}_{\gamma}t_Q}{4\pi n_{\rm HI}}\right) ^{1/3},
\label{rh2}
\end{equation}
where $n_{\rm HI}$ is the neutral hydrogen number density. Eq.(\ref{rh2}) applies for a homogeneous IGM and does not take into account both recombinations and relativistic effects. \\
The recombination timescale $t_{rec}$ is given by:
\begin{equation}
t_{rec}=\left[C\alpha_Bn_{\rm H}(1-x_{\rm HI})\right]^{-1},
\end{equation}
where $C\simeq 26.2917exp[-0.1822z+0.003505z^2]$ is the clumping factor \cite{Iliev07}, $\alpha_B=2.6\times 10^{-13}\rm{cm^3~s^{-1}}$ is the case B hydrogen recombination coefficient evaluated at $T=10^4~\rm{K}$, and $n_{\rm H}=7\times10^{-5}[(1+z)/7]^3 \rm{cm^{-3}}$ is the mean hydrogen number density. Thus, at the redshifts of interest $z\approx 6$, even in the limiting case $x_{\rm HI}=0$, $t_{rec}\approx 2\times 10^8 {\rm yr}$, thus being larger than typical QSOs lifetime $t_Q\approx 10^7$. This shows that eq.(\ref{rh2}) provides a plausible value for the HII region extent. For instance, Maselli et al. (2007) have shown that eq.(\ref{rh2}) matches quite well the mean value of the HII region size determined through radiative transfer calculations. \\
In this Section, we neglect relativistic effects which could squash the ionization front along the sight-line (\citeNP{white03}; \citeNP{wlb05}; \citeNP{Yu05}; \citeNP{shapirorel06}), possibly reducing the length of the lines of sight interested by the proximity effect. We will discuss this issue in detail in Sec. 5, when addressing the first observed case of transverse proximity effect.

At $z=5.5$, assuming 
$x_{\rm HI}=5.6\times 10^{-5}$ (see Fig. 1), $R_{HII}=1$ ($10$) Mpc could be produced by a 
QSO emitting a number of ionizing photons ${N}_{\gamma}=\dot{N}_{\gamma}t_Q=7\times 10^{65}$ ($7\times 10^{68}$). 
Thus, assuming a QSO lifetime $\approx 10^7 {\rm yr}$, the observed peaks in the LR (HR) sample 
require $\dot{N}_{\gamma}= 2.2\times 10^{51}$ ($2.2\times 10^{54}$) ${\rm s^{-1}}$, 
which would correspond to sources $\approx 6$ ($3$) orders of magnitude fainter than QSOs observed at 
$z\approx 6$, typically having $\dot{N}_{\gamma}\approx 10^{57} {\rm s^{-1}}$ \cite{haicen} and 
black hole masses $M_{BH}\approx 10^9{\rm M}_{\odot}$. 

So far we have assumed that the gas inside the HII region is fully ionized or, stated differently, that 
along the redshift path encompassed by the ionized bubble the flux is {\it completely} transmitted. 
However, this is unlikely since a sufficiently high opacity due to resonant (damping wing) optical depth associated with the neutral hydrogen 
inside (outside) the HII region can produce dark gaps. Thus, the relation between $P_{max}$ and $R_{HII}$ is
\begin{equation}
P_{max}=\frac{H(\bar{z})\lambda_{Ly\alpha}}{c}\frac{(1+\bar{z})}{(1+z_{em})} f_{t}R_{HII} = A(z) f_t R_{HII},
\label{ftransm}
\end{equation}
where $f_{t}$ is the mean transmitted flux computed inside the proximity region.  
We will derive $f_{t}$ in Sec. 5 from an observed case of transverse proximity effect, note
that values of $f_{t}<1$ would result in a larger luminosity of the QSO producing the 
transmissivity window.\\
Finally, powerful QSOs, as those observed at $z\approx 6$, could produce 
transmission windows consistent with observational data if they are embedded 
in overdense regions where the high density sustains an initial neutral 
fraction, $x_{\rm HI}\gtrsim 0.1$, {\it before the QSO turns on}. The expansion of 
the HII region in such environment would result in considerably smaller sizes 
(\citeNP{antiesimo}). In this case, both the host dark matter halo mass found 
above ($M_h\approx 10^{12}{\rm M}_{\odot}$), and the size of the HII region 
would combine to give the correct frequency and spectral width of the observed 
peaks.
\subsection{Galaxy HII regions}
In addition to QSOs, transmissivity windows could be produced by HII regions around high-$z$ galaxies. 
Adopting the canonical relations
\begin{equation}
M_*=f_*\frac{\Omega_b}{\Omega_m}M_h;\\
N_{\gamma}=\bar{n}_{\gamma}\frac{M_*}{m_p};\\
f_{esc}N_{\gamma}=\frac{4\pi}{3}n_{\rm HI}R_{HII}^3,
\end{equation}
where $M_*$ is the stellar mass, $\bar{n}_{\gamma}$ is the number of ionizing 
photons per baryon into stars, and $m_p$ is the proton mass, 
the relation between $M_h$ and $R_{HII}$ is given by:
\begin{equation}
M_h=3\times 10^8 M_\odot\left(\frac{1+z}{6.5}\right)^3y_{-1}R_{HII}^3,
\label{hm}
\end{equation}
where $y_{-1}=(x_{\rm HI} f_*^{-1}f_{esc}^{-1})/0.1$ and we are  
assuming $\bar{n}_{\gamma}=4000$, appropriate for a PopII stellar population with
a standard Salpeter IMF; we assume the fiducial values $x_{\rm HI}=5.6\times 10^{-5}$, $f_*=0.1$, 
$f_{esc}=0.01$.  The mass of an halo hosting a star-forming region able to produce 
$R_{HII}\approx 1$ ($10$) Mpc is $2\times 10^8$ ($2\times 10^{11}$) ${\rm M}_{\odot}$. 
At $z\approx 5.5$ objects of these masses corresponds to fluctuations of the 
density field $\gtrsim 1-\sigma$ ($2-\sigma$) \cite{baloeb}.\\ 
As for QSOs, the bubble size$-$peak frequency tension could be alleviated if 
the galaxies
live in overdense environments where the photoionization rate only supports a $x_{\rm HI}\approx 0.1$ 
(resulting in a larger value of $y_{-1}$, and hence of $M_h$ in eq.(\ref{hm})) prior to the 
onset of star formation in the galaxy.  Obviously, the previous arguments neglect that because
of clustering (\citeNP{yulu}; \citeNP{kramerhaiman}), as multiple sources could power a single HII region; 
in order to get firmer results radiative transfer cosmological simulations are required. 
\section{PEAKS FROM THE PROXIMITY EFFECT}
\begin{figure}
\psfig{figure=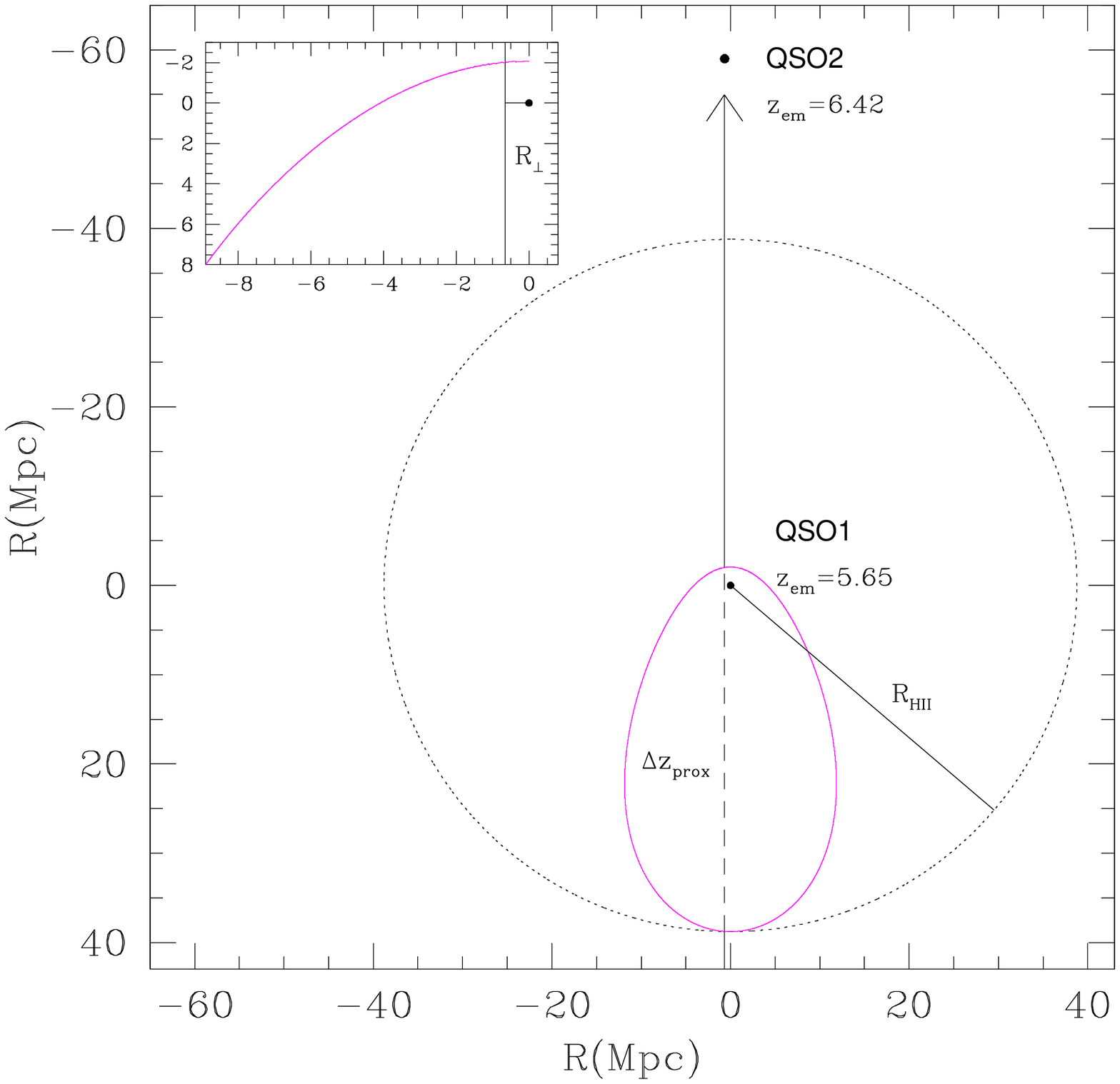,width=8.5cm,angle=0}
\caption{Schematic positions of quasars SDSS J1148+5251 
(QSO2, $z_{em}=6.42$, redshift position not in scale) and RD J1148+5252 (QSO1, $z_{em}=5.65$). The projected 
separation is denoted by $R_{\bot}$, the size of the HII region, $R_{HII}$, in the QSO1 rest frame is represented by the dotted circle; 
the magenta solid line shows the apparent shape of the ionization front; the dashed black line shows the redshift path ($\Delta z_{prox}$) 
in which the bubble produced by 
QSO1 intersects the LOS to QSO2. }
\label{prox}
\end{figure}

In Sec. 4, we have discussed the possibility that the observed peaks are 
produced by ionizing sources whose bubbles intersect the lines of sight to 
the target QSO. In this case one could ask if the source responsible for the 
HII region would be detected in the observed field. If the origin of 
transmissivity regions resides in bubbles around high-$z$ galaxies, these 
sources are too faint to be seen in the SDSS; however, deep HST 
imaging (\citeNP{stiavelli05}) could detect such objects. On the contrary, if 
the HII region of a quasar intervenes along the LOS to an 
higher redshift quasar, the first could be observed in the 
SDSS field. \\
\citeNP{mahabal05} have discovered a faint quasar 
(RD J1148+5253, hereafter QSO1) at $z=5.70$ in the field of the highest 
redshift quasar currently known (SDSS J1148+5251, hereafter QSO2) at $z=6.42$. 
In this Section we study the QSO2 transmitted flux, 
in order to analyze the proximity effect of QSO1 on the QSO2 spectrum. 
For clarity, Fig. \ref{prox} presents a schematic picture of the considered geometry.      
As the redshift $z=5.70$ quoted by \citeNP{mahabal05} is based on the peak of the 
Ly$\alpha$ emission line, and the estimated error from such procedure is $\Delta z\approx 0.05$ \cite{goodrich}, we assume $z_{em}^{QSO1}=5.65$ and we discuss this issue in further details in Appendix B.

The two QSOs have a projected separation of 109'', which corresponds to 
$R_{\bot}=0.66$ Mpc. The line of sight to QSO2 intersects the bubble produced 
by QSO1 for a redshift path ($\Delta z_{prox}$) whose length depends on the 
radius of the HII region ($R_{HII}$) itself. We find $R_{HII}=39~{\rm Mpc}$, 
by plugging in eq.(\ref{rh2}) the following values: 
$t_Q=1.34\times 10^7{\rm yr}$, $x_{\rm HI}=8.4\times 10^{-5}$, 
$\dot{N}_{\gamma}=8.6\times 10^{55}{\rm sec^{-1}}$, where $x_{\rm HI}$ is 
provided by the mean value between those predicted by our models 
at $z=5.7$ (see rightmost panel of Fig. \ref{prop}), while 
$\dot{N}_{\gamma}$ is compatible with the luminosity of a QSO 3.5 magnitudes 
fainter than QSO2 (\citeNP{mahabal05}). 
In this Section, we also take into account relativistic effects which could squash the ionization front along the LOS (\citeNP{white03}; \citeNP{wlb05}; \citeNP{Yu05}; \citeNP{shapirorel06}). The apparent size of the HII region, computed following the method outlined in \citeNP{Yu05}, is shown in Fig. \ref{prox}. By zooming the region close to QSO1 (small box in Fig. (\ref{prox})) it is clear that the apparent size of the HII region extends up to 2~Mpc in the 
direction toward QSO2.  
Given $R_{HII}$, taking into account relativistic effects, the region $\Delta z_{prox}$ extends from $z=5.16$ up to 
$z=5.68$. 
We re-compute $x_{\rm HI}$ along the LOS to QSO2, adding
to the 
uniform UVB photoionization rate $\Gamma_{\rm HI}$ the photoionization 
rate $\Gamma_{\rm HI}^{QSO1}$ provided by QSO1, obtained starting from the following equations:

\begin{equation}
\Gamma_{\rm HI}^{QSO1}=4\pi\int_{\nu_{\rm HI}}^{\infty}\frac{J_{\nu}}{h\nu}\sigma_0\left (\frac{\nu}{\nu_{\rm HI}}\right )^{-3}d\nu;
\end{equation}

\begin{equation}
J_{\nu}=\frac{\dot{N}_{\nu}h\nu}{16\pi^2 R^2};
\end{equation}

\begin{equation}
\dot{N}_{\gamma}=\int_{\nu_{\rm HI}}^{\infty}\dot{N}_{\nu}d\nu=\int_{\nu_{\rm HI}}^{\infty}\dot{N}_{\nu_{\rm HI}}\left(\frac{\nu}{\nu_{\rm HI}}\right)^{-{\alpha}}d\nu,
\label{ndot}
\end{equation}
where $\nu_{\rm HI}$ is the hydrogen photoionization 
frequency threshold, $\sigma_0$ is the Thompson scattering cross section, $R$ is the distance from QSO1 to the LOS, $\dot{N}_{\nu_{\rm HI}}$ is the rate of the emitted ionizing photons at the hydrogen photoionization frequency threshold and $\alpha=1.5$ is the spectral index of the QSO continuum.
Integrating eq.(\ref{ndot}) we obtain :
\begin{equation}
\dot{N}_{\nu_{\rm HI}}=\frac{(\alpha-1)\dot{N}_{\gamma}}{\nu_{\rm HI}}.
\end{equation}
Thus, it results:
\begin{equation}
\Gamma_{\rm HI}^{QSO1}=\left(\frac{\alpha-1}{\alpha+2}\right)\frac{\dot{N}_{\gamma}\sigma_0}{4\pi R^2},
\end{equation}

\begin{figure}
\psfig{figure=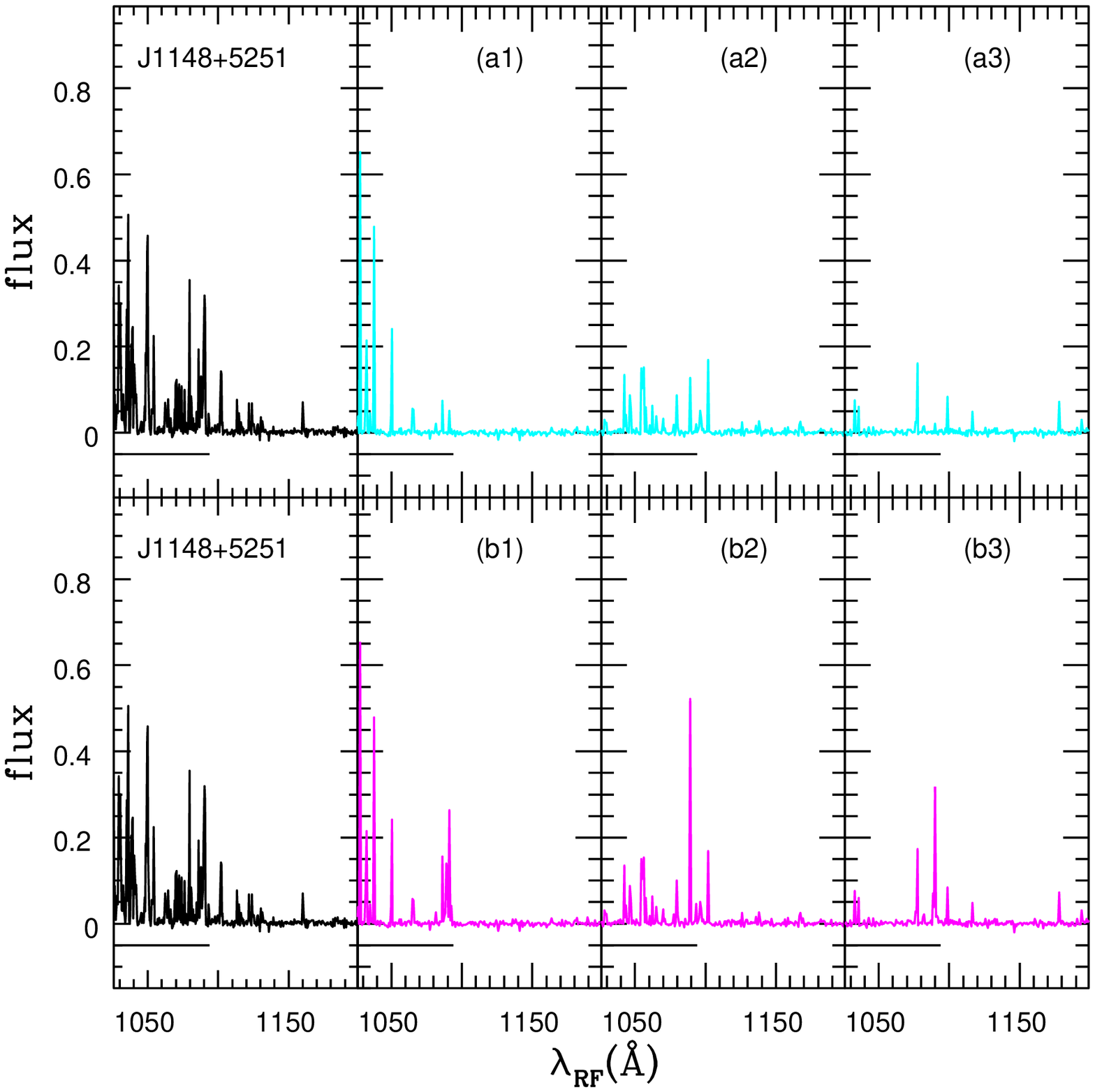,width=8.5cm,angle=0}
\caption{{\it Leftmost panels}: Observed transmitted flux (black spectra) 
in the spectrum of QSO SDSS J1148+5251 (QSO2, $z_{em}=6.42$). The solid black 
line shows the redshift path ($\Delta z_{prox}$) 
in which the bubble produced by 
QSO RD J1148+5252 (QSO1, $z_{em}=5.65$) intersects the LOS to QSO2.    
{\it Top panels (ai), with i=1,3}: 
Simulated fluxes (cyan spectra) along 3 different random LOS 
(cases ``without bubble''). {\it Bottom panels (bi), with i=1,3}: 
Simulated fluxes (magenta spectra) along the same LOSs 
shown in the top panels, taking into account the contribution from QSO1 to the 
total ionizing flux (cases ``with bubble'').}
\label{bubblemod}
\end{figure}

In Fig. \ref{bubblemod} we compare the observed transmitted flux in the spectrum of QSO2 with the simulated fluxes along 
3 different LOS with (bottom row) or without (top) including the contribution from QSO1 to the total ionizing flux. 
For brevity, we refer to these case as ``with bubble'' or ``without bubble''. Visual inspection of Fig. \ref{bubblemod} 
shows that the case ``with bubble'' is in better agreement with observations. Such statement can be made more 
quantitative by introducing a quantity denoted Peak Spectral Density (PSD), i.e. the number of peaks per
unit $\lambda_{RF}$ interval.
To compute the PSD for the two cases, we fix two different values for the flux threshold inside ($F_{th}^{IN}=0.01$) and outside 
($F_{th}^{OUT}=0.08$) the bubble. While $F_{th}^{OUT}$ is the same as the value used in this work so far, $F_{th}^{IN}$ has 
been chosen accordingly to the maximum observed noise r.m.s. deviation (for reasons explained in the Appendix A) in the $\lambda_{RF}$ 
interval $\Delta \lambda=1087-1092$~\AA, where $\Gamma_{\rm HI}^{QSO1}\gtrsim\Gamma_{\rm HI}$. For both the observed and simulated spectra, we compute the PSD 
inside and outside the bubble, finding the following results:
\\
\\
\centerline{
$(PSD_{obs}^{OUT},PSD_{obs}^{IN})=(0.11,0.40)$;
}
\\
\\
\centerline{
$(PSD_{sim}^{OUT},PSD_{sim}^{IN})=(0.04^{+0.06}_{-0.04},0.24^{+0.35}_{-0.24}),$
} 
\\
\\
where error bars provide the maximum and minimum PSD values found in the simulated LOS. 
Observationally, the PSD is found to be  
$\approx 4$ times\footnote{This factor depends on the flux threshold used. 
For example, it is reduced to $\approx 2.7$ if $F_{th}^{IN}=F_{th}^{OUT}=0.05$. 
Nevertheless, for the purpose of our test what really matters is the boost of 
this factor moving from outside toward inside the bubble.}
larger inside that bubble than outside it. This boost is quite well reproduced by the 
simulated PSD, although their absolute values are somewhat lower than the observed ones.

The physical interpretation of the results reported in this Section is the following. In 
the $\lambda_{RF}$ (redshift) interval $\Delta \lambda = 1087-1092$ ($\Delta z_{det} = 5.63-5.67$), where 
$\Gamma_{\rm HI}^{QSO1}\gtrsim\Gamma_{\rm HI}$, most of the gaps present in 
the case ``without bubble'' disappear, making room for peaks, 
as a consequence of the decreased opacity in the proximity of QSO1. Note that 
the $\Gamma_{\rm HI}$ value adopted in our calculations (shown in Fig.\ref{prop}, middle panel, red line) is close to the maximum value suggested by previous studies. 
Moreover, $\Gamma_{\rm HI}^{QSO1}\approx 3.4\times\Gamma_{\rm HI}$ at 0.66 Mpc from the foreground QSO. Thus, an implausible $\Gamma_{\rm HI}$ value should be assumed to explain the observed boost in the PSD with a uniform UVB. 
The enhancement in the transmissivity decreases for $\lambda_{RF}$ 
smaller (larger) than $1087$ ($1092$) \AA, since at the corresponding 
redshift $\Gamma_{\rm HI}^{QSO1}\lesssim \Gamma_{\rm HI}$. These results (i) confirm the detection
of a proximity effect, (ii)  show that the redshift stretch affected by the proximity effect 
is $ \Delta z_{det} <\Delta z_{prox}$.

\begin{figure}
\psfig{figure=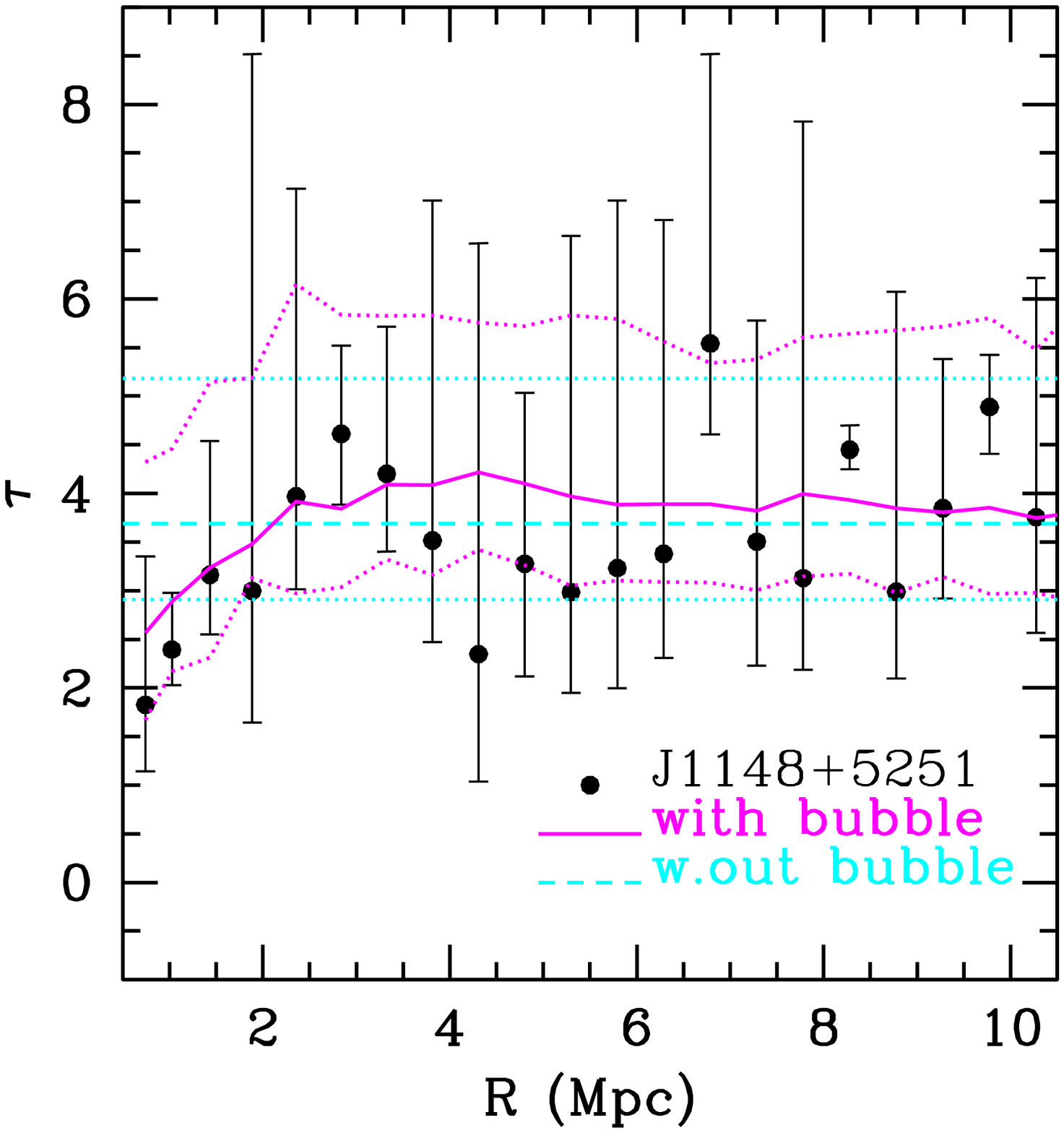,width=8.5cm,angle=0}
\caption{Evolution of the optical depth $\tau$ as a function of the distance R 
from QSO1. Filled circles denote the observed mean value for $\tau$, 
while error bars represent the maximum and the minimum observed $\tau$ at a 
given distance from the foreground QSO. Solid (dotted) magenta lines are the 
mean (maximum/minimum) values from 500 simulated LOS, computed adopting the 
case ``with bubble''. The dashed cyan horizontal line shows the mean optical 
depth predicted by the ERM in correspondence of the emission redshift of the 
foreground QSO. The dotted cyan horizontal lines denote the maximum/minimum 
optical depth at the same redshift.}
\label{overd}
\end{figure} 
As a final test for our model, we compute the observed evolution of the optical depth 
as a function of the distance $R$ from QSO1 and compare it with the predictions of model "with bubble'';
the result is shown in Fig. \ref{overd}. The agreement between observations 
and simulations is at 1-$\sigma$ confidence level for $86\%$ of the plotted 
points. For $R\lesssim 2$ Mpc, the mean optical depth $1.5 \lesssim\bar{\tau}\lesssim 3.5$ is 
lower than the mean value expected at $\bar{z}=5.65$ 
($\bar{\tau}_{5.65}\approx 4$); it approaches $\bar{\tau}_{5.65}$ at distances larger than $R_{\tau}\sim 2$ Mpc. \\
By taking the difference between $R_{\tau}$ and $R_{\bot}$, we set a lower limit on the foreground QSO lifetime $t_Q>\frac{R_{\tau}-R_{\bot}}{c}+(t_{\tau}-t_{QSO1})\approx 11$ Myr, where $t_{\tau}$ and $t_{QSO1}$ represent the cosmic times corresponding to the redshifts $z_{\tau}=5.68$ and $z_{em}^{QSO1}=5.65$, respectively.\\
It is worth noting that our model does not take into account neither (i) the clustering of the ionizing 
sources, nor (ii) the overdense environment expected around the QSO. Both these effects, 
in principle, could strongly affect the IGM ionization state, albeit in opposite ways.
While clustering of sources would enhance the transmissivity in the QSO near-zones, the 
overdense environment would tend to suppress it. The fact that we found agreement between 
observations and our modeling could indicate that, at least along this LOS, the two effects 
compensate. For what concerns (ii), by comparing the optical depth evolution observed 
in the proximity regions of 45 QSOs at $z_{em}\gtrsim 4$ with theoretical expectations, 
\citeNP{guima} find evidence for a density bias correlated with the QSO luminosity. 
Since QSO1 is much fainter than the QSOs studied by \citeNP{guima} it seems likely that 
neglecting such effect does not introduce a significant error. However, the extension of the 
proposed approach to a larger sample could clarify the relation between the clustering of sources 
and the overdensities in which massive objects are likely to be embedded.\\
It is important to note that the LOS toward SDSSJ1148+5251 contributes to the LPW distribution with the 
smallest peak ($P_{max}\approx 2$\AA) in the entire sample. Thus, even if we 
succeeded in reproducing the features of this LOS with our model 
``with bubble'', we still have to explain the mysterious origin of 
transmissivity windows as large as $10-15$\AA.\\
In Sec. 4.1, we estimate the QSO1 luminosity required to explain the 
observed $P_{max}$ value, and we comment on the result dependence from 
$f_{t}$. Plugging in eq.(\ref{ftransm}) the value $f_{t}\approx 0.03$
computed inside the proximity region, we obtain an effective size for $R_{HII}$; by 
further using eq.(\ref{rh2}), this translates into $\dot{N}_{\gamma}=9.2\times 10^{55}$~s$^{-1}$, 
a value in quite good agreement with the QSO1 ionizing rate quoted by \citeNP{mahabal05}.\\
\section{DISCUSSION}
We have studied several statistical properties of the transmitted flux in high-$z$ QSO spectra and
compared them with those obtained from simulated Ly$\alpha$ forest spectra to infer
constraints on the ionization state of the IGM at $z\approx 6$. We have considered 
two different reionization models: (i) an Early Reionization Model (ERM), 
in which the universe reionizes at $z_{rei}=6$, and (ii) a Late Reionization Model 
($z_{rei}\approx 7$).\\
By first using standard control statistics (mean transmitted flux evolution, probability 
distribution function of the transmitted flux, gap width distribution) in the
redshift range $3.5 < z < 6$, we show that both ERM and LRM match the 
observational data. This implies that current observations do not exclude that reionization can
have taken place at redshift well beyond six. \\ 
We then apply the Largest Gap Width (LGW) and Largest Peak Width  (LPW) statistics introduced 
by Gallerani et al. (2006) to a sample of 17 QSOs in the redshift range $5.74-6.42$. 
Both ERM and LRM provide good fits to the observed LGW distribution, favoring a scenario
in which $x_{\rm HI}$ smoothly evolves from $10^{-4.4}$ at $z\approx 5.3$ to $10^{-4.2}$ at $z\approx 5.6$. 

Discriminating among the two reionization scenarios would require
a sample of QSO at even higher redshifts. In fact, although according to LRM at $z\gtrsim 6$ the 
reionization process is still in the overlap phase with a mixture of ionized and neutral regions
characterizing the IGM, only $\approx 10\%$ of the simulated LOS pierce the overlap epoch,
and for a redshift depth $\Delta z \lesssim 0.2$. This explains why the predicted LGW 
distributions are quite similar for the two models considered.\\
Nonetheless, ERM provides a slightly better fit to observational data with respect to LRM, favoring 
$z_{rei}\gtrsim 7$. Within the statistical relevance of our sample, we have shown that LRM 
models can be used to put a robust upper limit $x_{\rm HI} < 0.36$ at $z=6.3$.\\

We have suggested that peaks preferentially arise from underdense regions of the cosmic density 
field and also from isolated HII regions produced by either faint quasars or galaxies. 
The frequency of the observed peaks implies that the dark matter halos
hosting such sources is relatively large, $\approx 10^{12}$ ($10^{13}$) $M_{\odot}$. 
Bright QSOs are unlikely to contribute significantly in terms of peaks, because 
given the required size of the HII regions, they should be located close enough to the LOS to 
the target QSO, that they should be detectable in the field.

The previous conclusions are substantiated by the specific case of an 
intervening HII region produced by the faint quasar RD J1148+5253 (QSO1) at 
$z=5.70$ along the LOS toward the highest redshift quasar currently known 
(SDSS J1148+5251, QSO2) at $z=6.42$.  It is worth noting that searches for 
the transverse proximity effect in the HI Ly$\alpha$ forest at $z\approx 3$ 
\cite{schirber} have been so far unsuccessful. Such effect has been isolated only by HeII absorption studies (\citeNP{WW06};\citeNP{WW07}). Thus, our results 
represent the first-ever detection in the HI Ly$\alpha$ forest.
We have analyzed the proximity effect of QSO1 on 
the QSO2 spectrum. Moreover, we have build up a simple model to estimate the 
location/extension of the proximity zone. Within the proximity region of QSO1
we have found an increased number of peaks per unit frequency with respect 
to segments of the LOS located outside the quasar HII bubble. This supports
the idea that we are indeed sampling the proximity region of the QSO1 and that
at least some peaks originate within ionized regions around (faint) sources. 
We then obtain a strong lower limit on the foreground QSO lifetime of $t_Q>11$ 
Myr.
Proper inclusion of galaxy clustering, which requires numerical simulations, 
might affect our conclusions \cite{FG07}.
Note that even in this clear-cut case, the size of the largest observed
peak in the spectrum of QSO2 is only of 2\AA.

Thus we are left with the puzzling discrepancy between observed and simulated transmissivity
windows (peaks) size, the former being systematically larger.  
Very likely, this reflects an unwarranted assumption made by the model. We do not believe that the discrepancy could be impute to the assumption
of a Log-Normal model, tested against \rm{HYDROPM} simulations by GCF06.
Nevertheless, we plan to compare the observed
Largest Peak Width
distribution with full hydrodynamical simulations in a future
work to study the correlation properties of the underdense regions, since Coles et al. (1993) have shown that the Log-Normal model produces a too ``clumpy'' distribution of the density field, when compared with N-body simulations.  

At least two physical effects, 
neglected here, could
affect the calculation of $x_{\rm HI}$: (i) non-equilibrium photoionization, and (ii)  
UV background radiation fluctuations. \\
The first assumption is made by the majority of studies dealing with the Ly$\alpha$ forest. However, if a fraction of the Ly$\alpha$ forest gas has been shock-heated as it condenses into the cosmic web filaments, it might cool faster than it recombines. For example, the recombination time $t_r$ becomes longer than the Hubble time when the density contrast is $\Delta < 7.5 [(1+z)/6.5]^{-3/2}$;
hence, large deviations from photoionization equilibrium are expected where 
$\Delta \ll 1$. Lower values of $x_{\rm HI}$ with respect to equilibrium are expected in such regions, as a result of the exceedingly slow recombination rates.\\

The second possible explanation for the too narrow simulated peaks might reside
in radiative transfer effects, also neglected here. At $z \approx 6$ the increase in the mean GP 
optical depth is accompanied by an evident enhancement of the dispersion of this measurement
which has been ascribed to spatial fluctuations of the UVB intensity near the end of reionization. 
A considerable (up to 10\%) scatter in the UVB HI photoionization rate  
is expected already at $z \approx 3$, as shown by \citeNP{antieandre} through
detailed radiative transfer calculations. The amplitude of such illumination 
fluctuations tend to increase with redshift because of the overall thickening of the forest. 
Although the observed dispersion in the mean GP optical depth may be compatible with a 
spatially uniform UVB (\citeNP{liu06}; \citeNP{LOF06}), it is likely that a proper radiative transfer treatment
becomes mandatory at earlier times. Basically, the main effect of fluctuations is to break 
the dependence of the HI neutral fraction on density. This is readily understood by considering 
two perturbations with the same density contrast $\Delta$. If the first is close to a luminous 
source it will have its  $x_{\rm HI}$ depressed well below that of the second one 
located away from it. Thus, opacity fluctuations naturally arise.
If so, peaks of larger width could be produced if the density perturbation associated with 
it happens to be located in a region where the UVB intensity is higher than the mean. 
\section*{Acknowledgments}
We thank R. Cen, B. Ciardi, D. Eisenstein, J.~P. Ostriker and S. White for 
stimulating discussions. We are particularly grateful to Z. Haiman, A. Lidz and A. Maselli 
for enlightening comments on the manuscript. 
XF acknowledges support from NSF grant AST 03-07384, a Sloan Research Fellowship, 
a Packard Fellowship for Science and Engineering. 
\bibliography{}
\bibliographystyle{mnras}
\section{APPENDIX A}
\begin{figure}
\psfig{figure=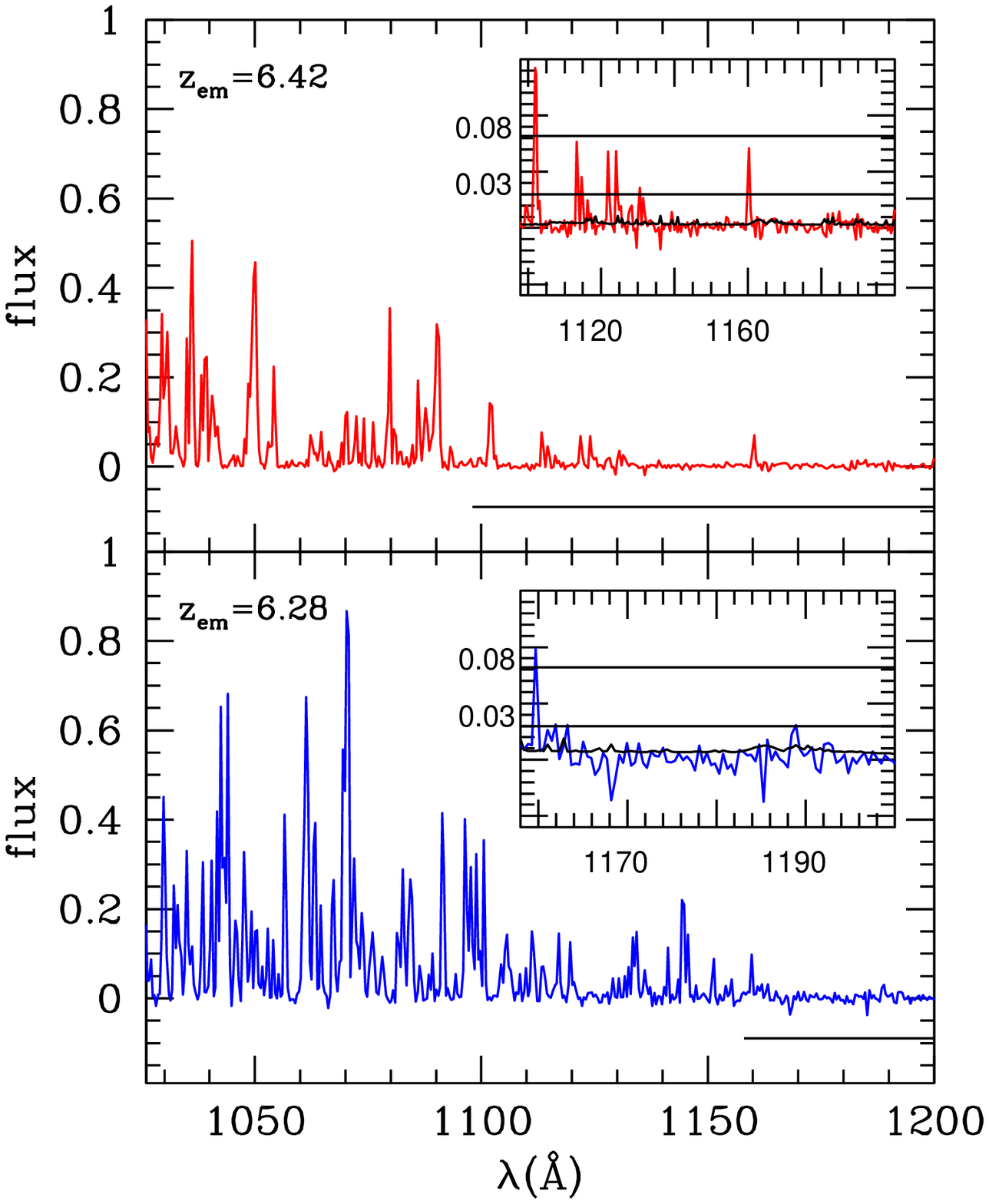,width=8.5cm,angle=0}
\caption{Observed spectra of the QSO SDSS J1148+5251 
({\it top panel}) and SDSS J1030+0524 ({\it bottom panel}). The 
black line denote the largest dark gap, measured by assuming $F_{th}=0.08$. 
In the small box the region interested by the largest dark gap is zoomed. The 
two black lines indicate $F_{th}=0.08$ and $F_{th}=0.03$. From the top 
(bottom) panel is evident that $F_{th}=0.08$ ($F_{th}=0.03$) overestimates 
(underestimates) the size of the largest dark gap.}
\label{2spc}
\end{figure}

The gap/peak statistics are sensitive to the S/N ratio, since spurious peaks 
could arise in spectral regions with noise higher than the flux threshold 
($F_{th}$) adopted. In what follows we restrict our attention to ``gaps'', 
since the extension of the conclusions on the ``peaks'' is direct. 
In particular, in this Appendix, we discuss the LGW distribution shape 
dependence on the $F_{th}$ chosen. 
We consider two values for $F_{th}$, namely $0.03$ and 
$0.08$, which correspond to $\tau=3.5$ and $\tau=2.5$, respectively. It is not 
obvious what criterion to apply in order to choose a proper value for 
$F_{th}$, since a too high (low) $F_{th}$ could overestimate (underestimate) 
the gap length. In Fig. \ref{2spc} we show two examples of 
spectra in which the $F_{th}$ choice strongly affects the gap measurement.
On the bottom, the spectrum of QSO J1030+0524 is shown, and, in the small box, 
the region marked by the solid black line is zoomed. It is evident that 
$F_{th}=0.03$ would break the gap at $\lambda_{RF}\approx 1190$ \AA, 
instead of at $\lambda_{RF}\approx 1160$\AA , 
as also noticed by F06; thus, in this case 
$F_{th}=0.08$ seems to be a better choice.\\
The opposite is true for the spectrum of QSO J1148+5251, shown in Fig. 
\ref{2spc} on the top. Indeed, in 
this case $F_{th}=0.08$ would provide a gap as large as $\approx 100$\AA, 
terminated by transmission at 
$\lambda_{RF}\approx 1100$\AA.  However, the peak at 
$\lambda_{RF}\approx 1160$\AA~is consistent with pure 
transmission (\citeNP{white03}; \citeNP{ohfurla}; F06); thus, in this case, 
$F_{th}=0.03$ would provide the correct gap measurement. For this reason, 
we compute the LGW distribution, considering both $F_{th}=0.03$ and 
$F_{th}=0.08$, alternatively. \\
The final LGW distribution is obtained as the mean of the 
preliminary ones, weighted on the corresponding errors. 
\section{APPENDIX B}
\begin{figure*}
\psfig{figure=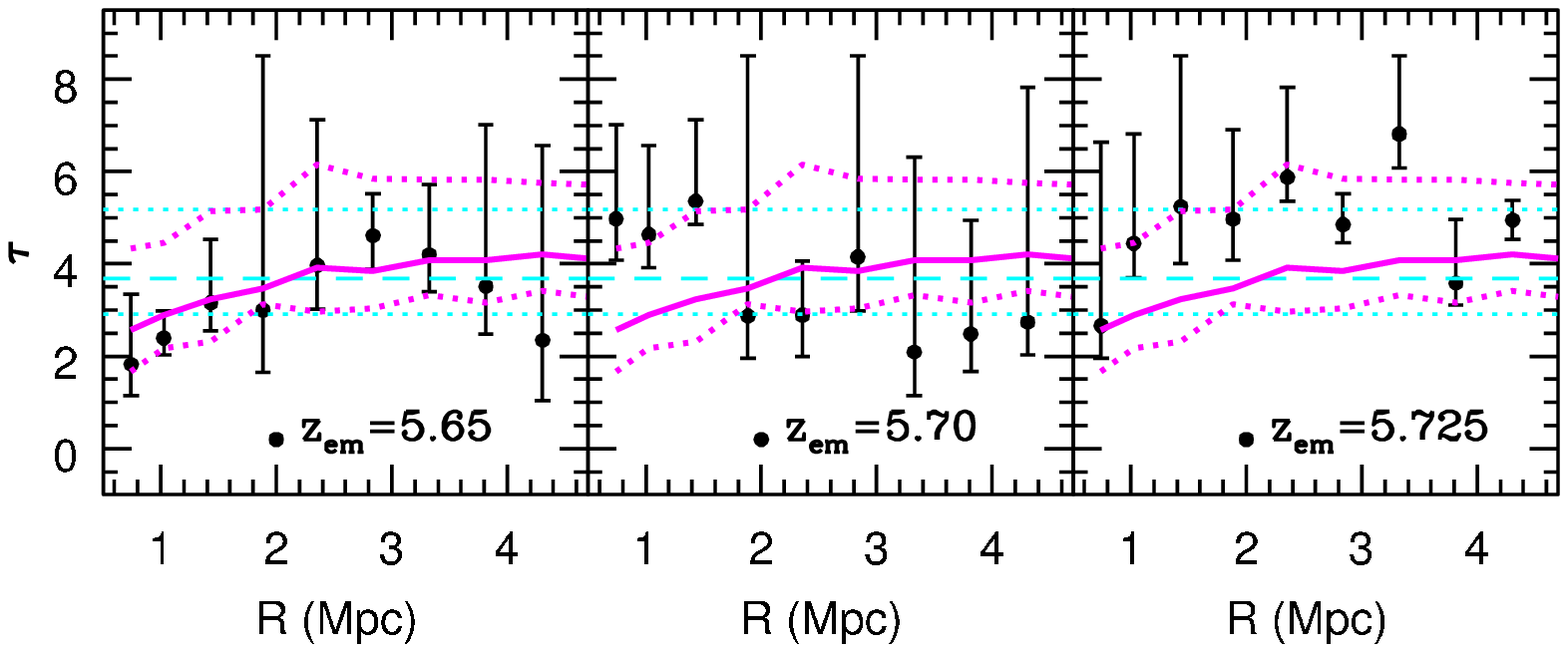,width=15.5cm,angle=0}
\caption{Same as Fig.7, but for different choices of foreground QSO emission redshift: $z_{em}=5.65$ (leftmost panel); $z_{em}=5.70$ (middle panel); $z_{em}=5.725$ (rightmost panel).} 
\label{compz}
\end{figure*} 

The redshift quoted by \citeNP{mahabal05} for RD J1148+5253 ($z_{em}=5.7$) is 
based on the peak of the Ly$\alpha$ emission line. This standard approach 
typically overestimates the true redshift by $\approx 0.05$ (e.g. \citeNP{goodrich}). For this reason 
we adopt as fiducial value $z_{em}=5.65$. 
As the object RD J1148+5253 (QSO1) is a BAL QSO \cite{mahabal05}, its emission 
redshift can not be established with accuracy from the broad metal lines, thus 
remaining uncertain.  
By comparing the QSO1 absorption spectrum with a BAL composite, 
also $z_{em}=5.725$ could be a plausible choice for the QSO1 emission 
redshift (Willot C., private communication). In this Appendix we repeat the 
analysis shown in Sec.5 considering different possibilities for the QSO1 
emission redshift. In Fig. \ref{compz} we 
compare the optical depth evolution as a function of the distance from QSO1 
obtained assuming $z_{em}^{QSO1}=5.65$ (left panel) with the cases in which 
$z_{em}^{QSO1}=5.70$ (middle panel) and $z_{em}^{QSO1}=5.725$ (right panel). In table \ref{tablepsd} the results of the Peak Spectral Density (PSD) for the tree different choices of QS01 emission redshift are shown, together with the wavelength interval $\Delta \lambda$ where $\Gamma_{\rm HI}^{QSO1}>\Gamma_{\rm HI}$.

\begin{center}
\begin{tabular}{|c|c|c|c|}
\hline
$z_{QSO1}$ & $PSD_{OUT}$ & $PSD_{IN}$ & $\Delta \lambda$\\ 
\hline
5.65 & 0.11 & 0.40 & 1087-1092\\ 
5.70 & 0.12 & 0.60 & 1095-1100\\ 
5.725 & 0.11 & 0.40 & 1099-1104\\ 
\hline
\label{tablepsd}
\end{tabular}
\end{center}
Even though Fig. \ref{compz} shows that the observed optical depth evolution 
as a function of the distance from QSO1 is better explained by our model 
assuming $z_{em}^{QSO1}=5.65$, this result does not rule out other choices of 
the QSO1 emission redshift. 
The results shown in table \ref{tablepsd} confirm the the evidence of the 
transverse proximity effect, since the boost in the PSD moving from outside 
toward inside the bubble is present in all the three cases considered. 
\end{document}